# Exploring trade-offs between landscape impact, land use and resource quality for onshore variable renewable energy: an application to Great Britain

R. McKenna[1,8], I. Mulalic[2], I. Soutar[3], J. M. Weinand[4], Price, J.[5], S. Petrović[1,6], K. Mainzer[7]

[1] Energy Systems Analysis, DTU Management, Technical University of Denmark, Denmark
[2] Department of Economics, Copenhagen Business School, Denmark
[3] Centre for Geography and Environmental Science, University of Exeter, UK
[4] Chair of Energy Economics, Karlsruhe Institute of Technology (KIT)
[5] UCL Energy Institute, UCL, UK
[6] Danish Energy Agency, Copenhagen, Denmark
[7] greenventory GmbH, Freiburg
[8] School of Engineering, University of Aberdeen, UK

**Abstract**

The ambitious Net Zero aspirations of Great Britain (GB) require massive and rapid developments of Variable Renewable Energy (VRE) technologies. GB possesses substantial resources for these technologies, but questions remain about which VRE should be exploited where. This study explores the trade-offs between landscape impact, land use competition and resource quality for onshore wind as well as ground- and roof-mounted photovoltaic (PV) systems for GB. These trade-offs constrain the technical and economic potentials for these technologies at the Local Authority level. Our approach combines techno-economic and geospatial analyses with crowd-sourced 'scenicness' data to quantify landscape aesthetics. Despite strong correlations between scenicness and planning application outcomes for onshore wind, no such relationship exists for ground-mounted PV. The innovative method for rooftop-PV assessment combines bottom-up analysis of four cities with a top-down approach at the national level. The results show large technical potentials that are strongly constrained by both landscape and land use aspects. This equates to about 1324 TWh of onshore wind, 153 TWh of rooftop PV and 1200-7093 TWh ground-mounted PV, depending on scenario. We conclude with five recommendations that focus around aligning energy and planning policies for VRE technologies across multiple scales and governance arenas.

# 1.    Introduction and overview

The United Kingdom (UK) has passed the 2008 Climate Change Act (UK Government, 2008), which sets a legally binding target to reduce indigenous greenhouse gas (GHG) emissions by 80% relative to 1990 levels by 2050. The Act was subsequently amended in 2019 to a 100% reduction target, i.e. net-zero GHG emissions, by 2050. The Climate Change Committee (CCC) has developed four pathways to demonstrate the ways in which the UK energy system can meet this target (CCC, 2020b). The four pathways have in common a further strong uptake of Variable Renewable Energy (VRE) technologies, especially solar photovoltaic (PV) as well as onshore and offshore wind – increasing from a total generation of 88 TWh in 2020 to 514 TWh in 2050.

Solar PV has seen strong growth in recent years, from just 0.04 TWh of electricity generation in 2010 to 13 TWh in 2019; of this, about 7 TWh is from systems larger than 5 MW and mainly ground-mounted and 5 TWh from smaller rooftop systems (BEIS, 2020b). Whilst new small systems below 5 MW are no longer eligible for Feed in Tariffs since 2019, existing plants continue to be supported in this way. Instead, in 2020 the Smart Export Guarantee (SEG) scheme was launched, which obliges licensed electricity suppliers (with over 150,000 customers) to offer small scale low-carbon generators a price per kWh of exported electricity (BEIS, 2020b). Current levelized costs of electricity (LCOEs) for these small-scale PV systems in a UK context are about 0.12 £/kWh (BEIS, 2020b), which is below the household electricity price of about 0.19 £/kWh (ofgem, 2021) and therefore economically incentivizes self-consumption.

Onshore wind will also undoubtedly play a crucial role in the future UK energy system, having increased from around 7 TWh of generation in 2010 to over 32 TWh in 2019 (BEIS, 2020b). This development was briefly slowed by a lack of political support for this technology, but the return to eligibility for the Contract for Difference (CfD) subsidies reverses this decision (DCLG, 2017). In addition, onshore wind has very high approval ratings amongst the public: a YouGov survey in 2018 found general support for onshore wind technology (YouGov, 2018). Overall support for renewable energy reached its highest ever level of 85% in 2018, increasing from 79% in 2017 (BEIS, 2018).

Despite this general approval, onshore wind proposals can encounter local opposition from planning authorities and local communities, especially if they are not directly engaged in the planning processes (Boudet, 2019; Fast et al., 2016). Visual impact is one of the central arguments from local residents against onshore wind installations (Molnarova et al., 2012; Petrova, 2016; Wolsink, 2018), although concern is reduced when people live further away from turbines (Betakova et al., 2015; Wolsink, 2018) and in contexts where the affected people have previous experience with wind energy (Schumacher et al., 2019; Sonnberger and Ruddat, 2017; van der Horst, 2007; Warren et al., 2005).

In order to reflect this lack of local support, studies devoted to resource assessments for onshore wind have recently tried to consider non-technical constraints and trade-offs between technologies (Höltinger et al., 2016; Jäger et al., 2016; McKenna et al., 2021c). But so far this has been focussed on wind, with little attention to ground-mounted PV systems.

The background analysis for the CCC study (Vivid economics and ICL, 2019) assessed feasible potentials for onshore wind and solar PV in the UK. Rather than consider all suitable locations and assess technical and economic feasibility of installing these technologies, this study pre-filtered the geographical potential[1] to exclude possibly unsuitable areas based on the quality of the wind resource. This means that the resulting potentials of 215-479 TWh for onshore wind and 35 TWh for rooftop PV (cf. Table 7) can be considered rather conservative (further comparisons with other studies are given in section 4.f).

The CCC scenarios are also strongly affected by land use competition between renewable energy technologies. This is a well-known and researched topic, especially but not only in relation to bioenergy and food. For example, Konadu et al. (2015) previously delineated this with respect to land use for bioenergy in a UK context. But land use competition was apparently not a focus of the (Vivid economics and ICL, 2019) study and has not been widely analysed for VRE technologies in the UK. Price et al. (2018) recently analysed how land and water restrictions can shape the least cost design of Great Britain's power system in 2050, but their analysis did not engage stakeholders or draw on empirical data to develop their land use constraints.

The present study takes this background as the starting point to analyse the potential future contributions of VRE technologies to the long term decarbonization target in the UK, in the context of the CCC (CCC, 2020b) pathways of 25-30 GW onshore wind and 75-90 GW solar in 2050[2]. We analyse and economically assess the technical potential for onshore wind, ground-mounted and rooftop PV with a detailed geospatial analysis of the whole of Great Britain (GB). We further explore the implications of aesthetic landscape impact and land-use competition on these potentials and costs within a quantitative framework. This paper thereby builds on and complements two related studies: firstly, in (McKenna et al., 2021c) we analysed the trade-off between scenicness and onshore wind costs at the national level; secondly, in Price et al. (2020) we explored the system-level impacts on the power system of this new dataset, for which the detailed methodology and spatially-disaggregated results are shown in the present paper. The key contributions and objectives of this paper are as follows:

1. Test the significance of any link(s) between scenicness data and VRE planning outcomes for both onshore wind and ground-mounted PV;
2. Provide a spatially disaggregated dataset of existing installed capacity and estimated resource potentials at Local Authority level across GB;

---

[1] Potential definitions can be found in McKenna et al. (2021b).
[2] With average annual full load hours from 2015-2019 of 2200 and 900, this equates to about 55-66 TWh and 68-81 TWh wind and solar respectively (BEIS (2020b).

3. Develop and apply a new combined top-down/bottom-up method for rooftop PV potentials at national scale;
4. Explore the impacts of scenicness and land competition for the three VRE technologies and derive insights relating energy and planning policy.

The paper is structured as follows. Section 2 introduces the employed scenicness dataset and establishes the statistical relationship between this and the outcome of planning applications for wind and solar plants. Section 3 presents the methodology for the resource assessments of onshore wind, ground- and rooftop PV. Section 4 then presents the results both at national and local levels, with a focus on the implications of scenicness and land-use competition. Section 5 is devoted to a discussion of the method and the results, and section 6 concludes with policy implications.

## 2. Scenicness and planning applications for renewable energy plants

This section presents the "Scenic or Not" dataset in section 2.a before analyzing the link between scenicness and planning applications for wind and PV in section 2.b.

### a. Scenic or Not dataset

Here we analyse the association between the scenicness and the planning outcome of energy projects using scenic ratings from *Scenic-Or-Not* (http://scenicornot.datasciencelab.co.uk/) as a measure of scenicness and detailed data about renewable energy applications in Great Britain from the Renewable Energy Planning Database (BEIS, 2014). Users of *Scenic-Or-Not* have rated random geotagged photographs taken at 1km$^2$ resolution for the whole of Great Britain on an integer scale of 1–10, where 10 indicates "very scenic" and 1 indicates "not scenic". The database contains 1,536,054 ratings for 212,212 images. We use the mean scenicness values for all photos rated three times or more taken at the locations after the energy project has been implemented.

The Renewable Energy Planning Database (BEIS, 2014) includes the date of the application, operator, information on the site, project attributes (e.g. technology and capacity), and the outcome of the application (granted or rejected) for plants larger than 150 kW. This database has previously been employed by Roddis et al. (2018) and Harper et al. (2019) in a similar manner, but without any scenicness data.

For onshore wind energy the mean success rate is about 0.6 (514 project applications have been rejected and 740 have been granted for the time period 2001-2017) while for the ground-mounted PV projects the mean success rate is about 0.8 (1,275 project applications have been granted and 282 have been rejected). Moreover, in order to account for highly-sensitive areas, we compute for all locations in our database the distance to the closest Special Areas of Conservation (SAC), distance to the closest Special Protection Areas (SPA), distance to the closest Ramsar areas (wetlands), distance to the closest National Park, and distance to the closest airport.

The results for onshore wind are taken from (McKenna et al., 2021c); here we apply the same method to explore ground-mounted PV systems. For more details on the data see (McKenna et al., 2021c).

### b. Logistic regression of planning applications for wind and solar

We assume a standard model specification for the planning outcome for a project application $i$ at year $t$:

$$\Pr(D_{i,t} = 1 \mid S, X; \alpha, \beta, \boldsymbol{\delta}, \boldsymbol{\gamma}) = F(\alpha + \beta\, S_{i,t} + \boldsymbol{\delta}' X_{i,t} + \boldsymbol{\gamma}_t) \quad (1)$$

where $D_{i,t}$ denotes the discrete dichotomous variable taking a value of 1 if the application decision is positive, otherwise 0; α is a constant term and γ is the year fixed effect; $S_{i,t}$ is the scenicness value; and $X_{i,t}$ denotes controls for project characteristics such as technical and geographical attributes. We assume that the error term is identically and independently Extreme Value type I distributed (i.i.d. EV I) and estimate the model coefficients using maximum likelihood, viz. logit regression (McFadden, 1974). We are particularly interested in the value of $\beta$, as if the scenicness is not related to the application decision then $\beta = 0$, whereas $\beta < 0$ if the scenicness value is negatively related to the planning outcome.

Table 1 shows the results of the logit. The upper panel (a) reports the estimation results for the wind energy projects and the lower panel (b) for the ground-mounted PV projects. Model 1 includes only the scenicness value and in the following models 2-4 we sequentially introduce the year fixed effects (to account for possible year-specific structural trends such as business cycles, inflation rate and political environment), the project size, and the environmental variables, respectively. For the wind energy projects the estimated odds ratios associated with the scenicness value are below one (estimated coefficient are negative) and highly significant. Model 4 is our preferable specification. This model suggests that for every one unit increase in the scenicness value, we expect a 0.22 decrease in the log-odds of a positive application decision, all else being equal. For the solar PV projects the estimated odds ratios associated with the scenicness value are close to one and never significant, suggesting that the impact of the ground-mounted solar panels on landscape aesthetics is less pronounced.

We have performed a number of sensitivity analyses. First we have assumed that the error term is i.i.d. normally distributed, viz. probit regression. The results of this probit regression model are very similar to the preferable model (4). We have also estimated logit models on a subsample when the number of votes is larger than 11 (10% percentile), 15 (25% quartile) and 25 (median), but the coefficients remain unchanged. The estimation results for these models are available upon request from the authors.

*Table 1: Logit regression results (odds-ratio) for onshore wind (McKenna et al., 2021c) and ground-mounted PV project planning outcomes*

|  | Model 1 | Model 2 | Model 3 | Model 4 |
|---|---|---|---|---|
| | (a) Wind energy projects | | | |
| **Scenicness value** | 0.844*** | 0.796*** | 0.768*** | 0.778*** |
| | (0.033) | (0.035) | (0.036) | (0.038) |
| **Number of turbines** | | | 1.225*** | 1.222*** |
| | | | (0.032) | (0.032) |
| **Capacity (MW)** | | | 0.936*** | 0.936*** |
| | | | (0.008) | (0.008) |
| **log distance to the closest National Park** | | | | 1.171*** |
| | | | | (0.068) |
| **log distance to the closest airport** | | | | 0.984 |
| | | | | (0.114) |
| **log distance to the closest Special Protection Areas (SPA)** | | | | 0.972 |
| | | | | (0.044) |
| **log distance to the closest Special Areas of Conservation (SAC)** | | | | 0.874** |
| | | | | (0.054) |
| **log distance to the closest Ramsar areas** | | | | 1.040 |
| | | | | (0.064) |
| **Year fixed effect** | no | yes | yes | yes |
| **Constant** | 2.930*** | 162.290*** | 133.568*** | 98.401*** |
| | (0.518) | (167.191) | (138.216) | (111.718) |
| **Number of observations** | 1,252 | 1,252 | 1,252 | 1,252 |
| **AIC** | 1,683 | 1,461.83 | 1,371.13 | 1,370.11 |
| **Log likelihood** | -839.40 | -717.92 | -846.94 | -665.05 |
| | (b) Ground-mounted PV projects | | | |
| **Scenicness value** | 0.973 | 0.972 | 0.972 | 0.970 |
| | (0.050) | (0.051) | (0.051) | (0.052) |
| **Capacity (MW)** | | | 0.987 | 0.987 |
| | | | (0.008) | (0.008) |
| **log distance to the closest National Park** | | | | 1.107* |
| | | | | (0.066) |
| **log distance to the closest airport** | | | | 1.232** |
| | | | | (0.111) |
| **log distance to the closest Special Protection Areas (SPA)** | | | | 0.971 |
| | | | | (0.093) |
| **log distance to the closest Special Areas of Conservation (SAC)** | | | | 0.755*** |
| | | | | (0.061) |
| **log distance to the closest Ramsar areas** | | | | 1.027 |
| | | | | (0.085) |
| **Year fixed effect** | no | yes | yes | yes |
| **Constant** | 4.985*** | 3.344** | 3.510** | 1.843 |
| | (1.004) | (1.857) | (1.952) | (1.257) |
| **Number of observations** | 1,558 | 1,558 | 1,558 | 1,558 |
| **AIC** | 1,480 | 1,434 | 1,434 | 1,423 |
| **Log likelihood** | -738.16 | -709.29 | -708.02 | -697.44 |

*Note: the dependent variable is a discrete dichotomous variable taking a value of 1 if the application decision is positive, otherwise 0; ***, **, * indicate that estimates are significantly different from zero at the 0.01, 0.05 and 0.10 levels, respectively; standard errors are in parentheses. AIC is Akaike's information criterion.*

## 3. Resource assessment method for VRE technologies

The method employed in this paper involves first determining the geographical potential, followed by the technical one, which is economically assessed (McKenna et al., 2021b). This procedure involves the stepwise removal of unsuitable (negative) areas from total available areas in a Geographical Information System (GIS), leaving suitable (positive) areas or polygons (cf. Figure 1). The standardized parts of the method are summarily reported with references to literature for details, and the focus in this section is on the new aspects of the methodology. This section presents the method for ground-mounted PV in section 3.a, rooftop PV in 3.b and onshore wind in 3.c. Section 3.d explains the LCOE calculation, 3.e is about matching multiple spatial datasets for VRE capacities in the UK and 3.f presents the analysed scenarios.

### a. Ground-mounted PV

For ground-mounted PV, the geographical potential is determined as follows. The maximum terrain slope that guarantees the technical feasibility of a ground-mounted PV plant is 15° (Borgogno Mondino et al., 2015; Carrión et al., 2008; Perpiña Castillo et al., 2016). Thus, all areas steeper than 15° are excluded using the OS Terrain 50 data (Ordnance Survey, 2018). All protected areas such as Ramsar areas, Special Areas of Conservation (SAC), Special Protection Areas and the National Parks are excluded. The National Parks data are extracted from the (Office for National Statistics, 2020) and the other protected areas are retrieved from the Joint Nature Conservation Committee (JNCC, 2016). Urban areas are extracted from Open Street Map (OpenStreetMap Contributors, 2020). High quality agricultural land is extracted from different websites depending on the country: for England, the data come from the Ministry of Agriculture, Fisheries and Food (1988), for Scotland, Scotland's soil (1981) and for Wales, it is the Welsh Government (Welsh Government, 2017). Agricultural lands of England and Wales are graded one to five with one the highest quality and five the poorest. The scale is different for Scotland; they are graded from one to seven with one being the highest quality and seven the poorest. Based on the description of each level of agricultural land for each country, the levels are matched into a single classification (from 1 to 5) across Great Britain. Thereby, the Scottish Classes 3 and 4 were equated with the Subgrades 3a and 3b in the English and Welsh classification, and Classes 6 and 7 were equated with Grade 5. In order to investigate the land use competition between the agriculture land (for food or bioenergy) and ground-mounted PV, two different types of restriction are considered: a low restriction in which the lands with grade 1 and 2 are excluded, and a high restriction in which lands with grade 3 are excluded as well (cf. scenarios in section 2.f).

The technical potential for ground-mounted PV, $E_{PVG}$, is determined with solar radiation data SAF "SARAH" from PVGIS (European Commission, 2018), which gives the yearly average global irradiance on a horizontal surface (W/km$^2$), H, for the whole of Great Britain, as long-term averages for 2005-2016. Multiplying by the hours in a year h (i.e. 8760) yields the annual irradiance in kWh/m$^2$. The data is converted from raster to vector in order to intersect with the geographical potential. Optimal inclinations for all parts of the world are taken from PV GIS (European Commission, 2019), for GB this varies from 30° in the south to 40° in the north. It is assumed that all modules are oriented south facing. Hence using (Mainzer et al., 2014) yields the relative solar irradiation on the inclined surface relative to a horizontal surface (inclination of 0°), meaning a maximum of 17% increase from the horizontal at 30-40°. The Packing Factor (PF) considered to account for space between modules is based on the median value from (Ong et al., 2013) of 51%. The final step is a Performance Ratio (PR) of 85% and efficiency η of 15% based on (Hoogwijk, 2004; Mainzer et al., 2014), which corresponds to polycrystalline silicon, the most dominant technology on the market (Gangopadhyay et al., 2013). Equation 2 defines the technical generation potential:

$$E_{PVG} = 117\% \cdot h \cdot \eta \cdot H \cdot A \cdot PR \cdot PF \qquad (2)$$

with A the total area [m2].

### b. Rooftop PV

For rooftop PV, the geographical potential is determined based on a combination of the bottom-up approach based on (Mainzer et al., 2017), and the top down method based on (Mainzer et al., 2014), with different data sources to transfer the method from a German to British context (see Gassar and Cha (2021) for a review of these methods). The bottom-up approach employs Bing Maps and Open-Street-Map alongside machine learning to recognize rooftop geometry and features. The method has a high resolution at the individual building level, but cannot be employed for the whole of the UK for computational reasons. Instead the four cities of Leeds, Glasgow, Birmingham and London are chosen based on expert discussions about the diversity of the building stock here (Liddiard, 2021). To connect the bottom-up method from (Mainzer et al., 2017) with the top-down approach, the following variables are employed: land area A is total area in m² in a specific land use category; building footprint area s is the plan outline of a building in m²; and roof area U is the total roof area in m² (if a flat roof, equal to the footprint area). Furthermore, a dimensionless ratio r is found for the four cities between the land area of each land use category i and the total footprint area of the buildings that fall into this land use type s. For this, the following land-use categories from CORINE land cover (EEA, 2012) at a resolution of 100 m for 2012 are employed (the update to 2018 was not yet available at the time this study): 111: continuous urban fabric; 112: discontinuous urban fabric, 121; industrial or commercial unit. The ratio r is calculated according to Equation 3.

$$r = \frac{s}{A} \qquad (3)$$

The assumption is made that, as 83% of the UK population lives in cities (DEFRA, 2019), that the majority of buildings and rooftop potentials are also in cities and therefore this method combining bottom-up and top-down approaches is highly transferable.

The usable roof area U is determined through the cosine of the footprint area, for each of the 72 azimuth/tilt combinations obtained from the method in (Mainzer et al., 2017). A utilization factor is not required as the bottom-up methodology already delivers partial (i.e. suitable) roof areas (Figure 8 shows some of these), excluding chimneys, obstructions etc. Equation 4 defines the useable roof area U per azimuth/tilt class i:

$$U_i = A \cdot r \cdot \frac{p_i}{\cos(v_i)} \quad (4)$$

where $p_i$ is the proportion of class i, and $v_i$ is the tilt of the roofs in class i.

The technical potential energy yield $E_{PVR}$ is determined in the same manner as for ground-mounted PV above, see Equation 5, whereby the yields are looked up in (Mainzer et al., 2014) and $irr_i$ is the relative irradiance for class i.

$$E_{PVR} = h \cdot \eta \cdot H \cdot PR \cdot \sum_{i=1}^{72} U_i \cdot irr_i \quad (5)$$

The LCOEs are calculated with the method in section 3.d below.

### c. Onshore wind

The approach to assessing the technical potential for onshore wind is very similar to one applied in (McKenna et al., 2014) and (McKenna et al., 2021c), with the results adopted from the latter source. The geographic potential is determined by excluding unsuitable areas for onshore wind, as follows. Based on Digital Elevation Models, regions with a slope over 20° are excluded, with slopes calculated in the same way as for ground-mounted PV (cf. section 3.a). Unsuitable areas are then excluded based on CORINE (EEA, 2012) data at 100 m from 2012. For this reason, the CORINE data is combined with OSM data to improve the coverage in and around urban areas. There is no general guidance on the buffer distances to housing and this differs between countries (McKenna et al., 2021b), but some guidance suggests 350 m for England, 2 km for Scotland and 500 m for Wales, which are the assumptions adopted (Barclay, 2012). Otherwise the assumptions relating to offset distances are based on Table 4.2 in (McKenna et al., 2014). The combination of the two datasets, OSM and CORINE, is based on the following precedence relationships. The basic approach is to prioritize OSM data where this is available, i.e. if both CORINE and OSM refer to the same polygon with (different) information, OSM takes precedence. This is based on the following steps, whereby positive and negative areas represent those suitable and unsuitable for wind energy respectively: 1. $OSM_{Positive} - CLC_{Positive}$; 2. $CLC_{Positive} - OSM_{Negative}$; 3. Unite (1) and (2); and 4. combine with OSM total area to obtain the land use categories (where available).

The technical potential for onshore wind energy is then calculated for the determined geographic potential area, again based on the method in (McKenna et al., 2014). This involves combining the land-use data with annual average wind speeds at 5 $km^2$ resolution and 10 m above ground for 2001 to 2006 inclusive from the UK Met Office (UK Met Office, 2018). These wind speeds are extrapolated to hub heights, and converted from average wind speeds to average wind speed distributions based on the roughness length of the land-use category (Silva et al., 2007). Finally, the energy generation for specific turbines is simulated based on the product of the integral of the wind speed distribution and the turbine's power curve. The most economical turbine for a specific location is determined by minimizing the LCOEs (section 3.d), based on a database of techno-economic turbine characteristics.

### d. Levelized Costs of Electricity (LCOE)

The Levelized Costs of Electricity (LCOEs) are calculated in the same way for each technology, based on the following Equation 6:

$$LCOE = \frac{I_0 \cdot \sum_{t=1}^{n} \frac{M_t}{(1+i)^t}}{\sum_{t=1}^{n} \frac{E_t}{(1+i)^t}} \quad (6)$$

where n is the lifetime of the technology, $I_o$ the investment [£], $M_t$ the annual costs in year t [£/year], $E_t$ energy produced in year t [MWh/year] and i the interest rate. Table 2 below gives an overview of the economic assumptions for the studied technologies. To reflect a private investor's perspective, i is assumed to be 8%.

*Table 2: Economic assumptions for VRE technologies (BEIS, 2020d; Open Energy Information, 2017; Philipps and Warmuth, 2019; Vartiainen et al., 2020)*

| Technology | Investment (£/kW) | O&M costs | O&M cost units | Lifetime (years) |
|---|---|---|---|---|
| Ground-mounted PV | 500 | 8.00 | (£/kW.year) | 20 |
| Rooftop PV | 1130 | 9.57 | (£/kW.year) | 20 |
| Onshore wind | 1050 | 0.02 | £/kWh | 20 |

### e. Matching multiple local datasets of VRE capacities

In order to underpin the disaggregated geospatial analysis carried out in this paper, multiple different databases relating to the installed capacities of VRE technologies were combined. The table "Renewable electricity by local authority" (BEIS, 2020c) contains all VRE generation on a Local Authority level for the whole of the UK. The Renewable Energy Planning Database (BEIS, 2014) mentioned above only includes larger VRE plants, i.e. those requiring planning permission and therefore over 150 kW rated capacity (BEIS, 2014). A further table, "Sub-national-total-final-energy consumption…" (BEIS, 2020e) includes disaggregated electricity demand figures at the LA level. Linking these databases is not trivial, for example because the first one contains nine-digit LA codes and the second contains X-Y coordinates and in some cases postcodes. These two databases were therefore linked on the basis of LA codes and the postcodes, by employing an online batch lookup tool (Ordnance Survey, 2021). The matched database is employed for the results analysis in section 4 and is provided as supplementary material to this article (McKenna et al., 2021a).

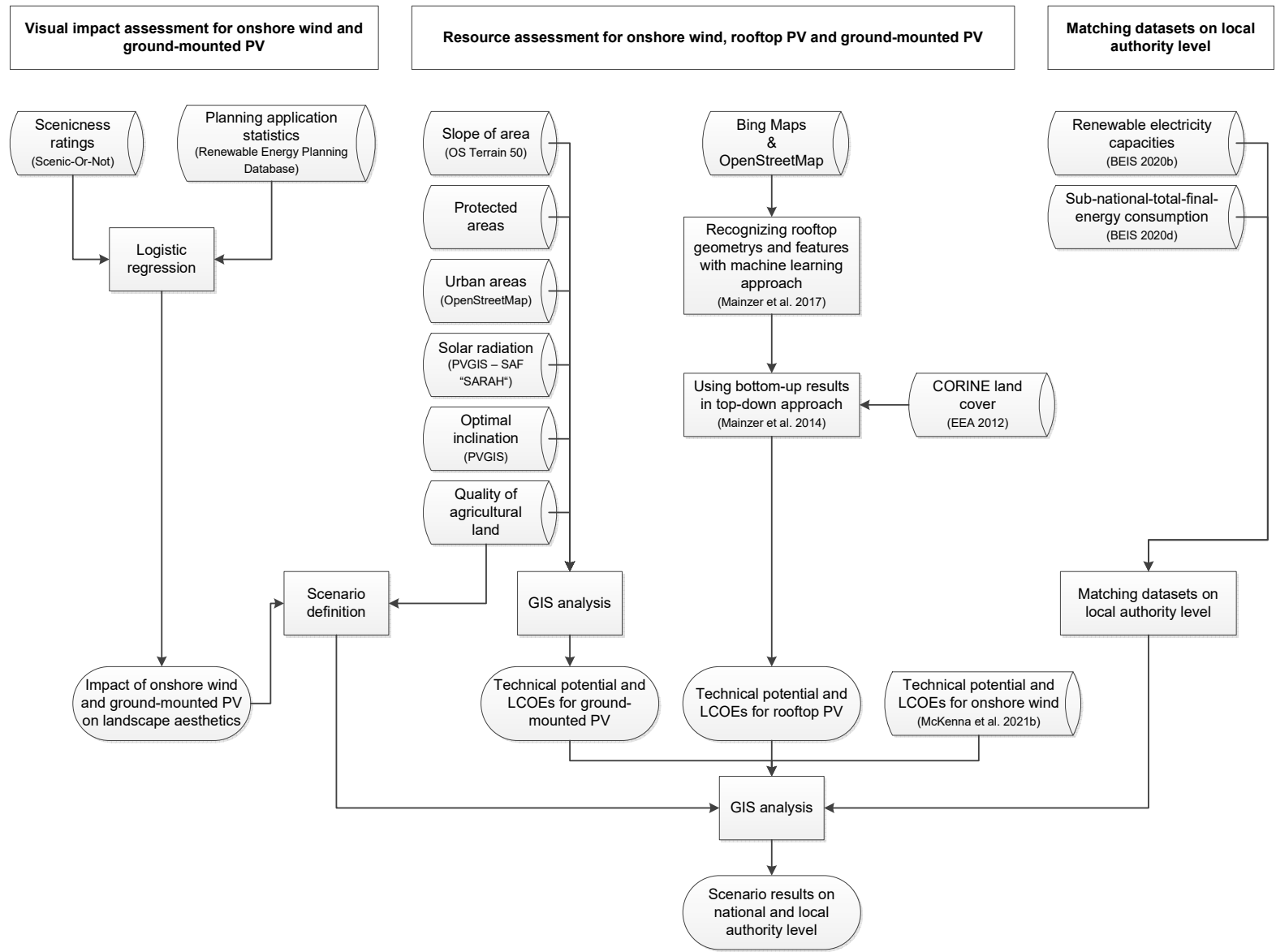

*Figure 1: Schematic of the approach of this paper, showing techno-economic and GIS analysis of 3 VRE technologies*

### f. Accounting for scenicness and definition of scenarios

In this paper we define eight high-level scenarios that are intended to explore the solution space under consideration. The result from section 2 above demonstrates that the scenicness only has a statistically significant correlation with the planning outcomes for onshore wind (and not ground-mounted PV). It is also well researched that the landscape visual impact of rooftop PV is minimal. In the German, Swiss and French regions of the Upper Rhine Region, for example, 75% of those surveyed stated that the distance to a rooftop PV system is not relevant for their acceptance, which also reflects the wider acceptance literature (Schumacher et al., 2019). Therefore we consider four scenarios for the onshore wind potential, based on gradually reducing the technical potential by quartiles of the scenicness distribution. The ground-mounted PV potential is delineated into two scenarios based on high and low restrictions to reflect agricultural land quality. Due to its limited interaction and land use competition with other technologies, the rooftop PV scenario merely reflects the technical potential. An overview of these eight scenarios is given in Table 3 below.

*Table 3: Overview of eight analysed scenarios for onshore wind, ground-mounted and rooftop PV potentials*

| Scenarios | Wind onshore | | Ground-mounted PV | | Rooftop PV | |
|---|---|---|---|---|---|---|
| | Rationale | Definition | Rationale | Definition | Rationale | Definition |
| 1 | Technical potential | Scenicness <=10 | Technical potential with high restriction | Only agricultural land categories 4-5 are feasible | Technical potential | All partial rooftop areas across 72 inclination and azimuth categories |
| 2 | 75% scenicness | Scenicness <=5.80 | | | | |
| 3 | 50% scenicness | Scenicness <=4.67 | | | | |
| 4 | 25% scenicness | Scenicness <=3.67 | | | | |
| 5 | Technical potential | Scenicness <=10 | Technical potential with low restriction | Only agricultural land categories 3-5 are feasible | | |
| 6 | 75% scenicness | Scenicness <=5.80 | | | | |
| 7 | 50% scenicness | Scenicness <=4.67 | | | | |
| 8 | 25% scenicness | Scenicness <=3.67 | | | | |

## 4. Results and discussion

This section presents the results, starting with high-level national results in the eight scenarios in section 4.a, followed by results at the Local Authority level in section 4.b. Section 4.c then explores the regional scenicness impacts on onshore wind before section 4.d analysis the land-use competition between areas for ground-mounted PV and onshore wind.

### a. Overall results in context

Overall the results show some large technical potentials for the three technologies analysed, as shown in Table 4 and Figure 2 below. This equates to about 1324 TWh of onshore wind, 153 TWh of rooftop PV and 1200-7093 TWh ground-mounted PV. These results relate to total areas of 11-80 million km$^2$, 1,190 km$^2$ and 15-93 million km$^2$ for onshore wind, rooftop and ground-mounted PV respectively. Consecutively removing the most-scenic locations based on quartiles of the distribution reduces the onshore wind potential to 962 TWh, 586 TWh and 267 TWh up to and including scenicness thresholds of 5.8, 4.67 and 3.67 respectively (McKenna et al., 2021b).

*Table 4: Overall results showing total available area and generation potential in the eight scenarios shown in Table 3*

| Scenarios | Rooftop PV area (km$^2$) | Rooftop PV (TWh) | Wind area (Mkm$^2$) | Wind potential (TWh) | Ground-mounted PV area (Mkm$^2$) | Ground-mounted PV potential (TWh) |
|---|---|---|---|---|---|---|
| 1 | 1190 | 153 | 81 | 1324 | 93 | 7093 |
| 2 | | | 35 | 962 | | |
| 3 | | | 22 | 586 | | |
| 4 | | | 11 | 267 | | |
| 5 | | | 81 | 1324 | 15 | 1051 |
| 6 | | | 35 | 962 | | |
| 7 | | | 22 | 586 | | |
| 8 | | | 11 | 267 | | |

The LCOEs for these technologies vary, from 0.12 £/kWh for rooftop PV, increasing rapidly to over 0.20 £/kWh, around 0.06 £/kWh to 0.10 £/kWh for ground-mounted PV and upwards of 0.04 £/kWh for onshore wind (Figure 2).

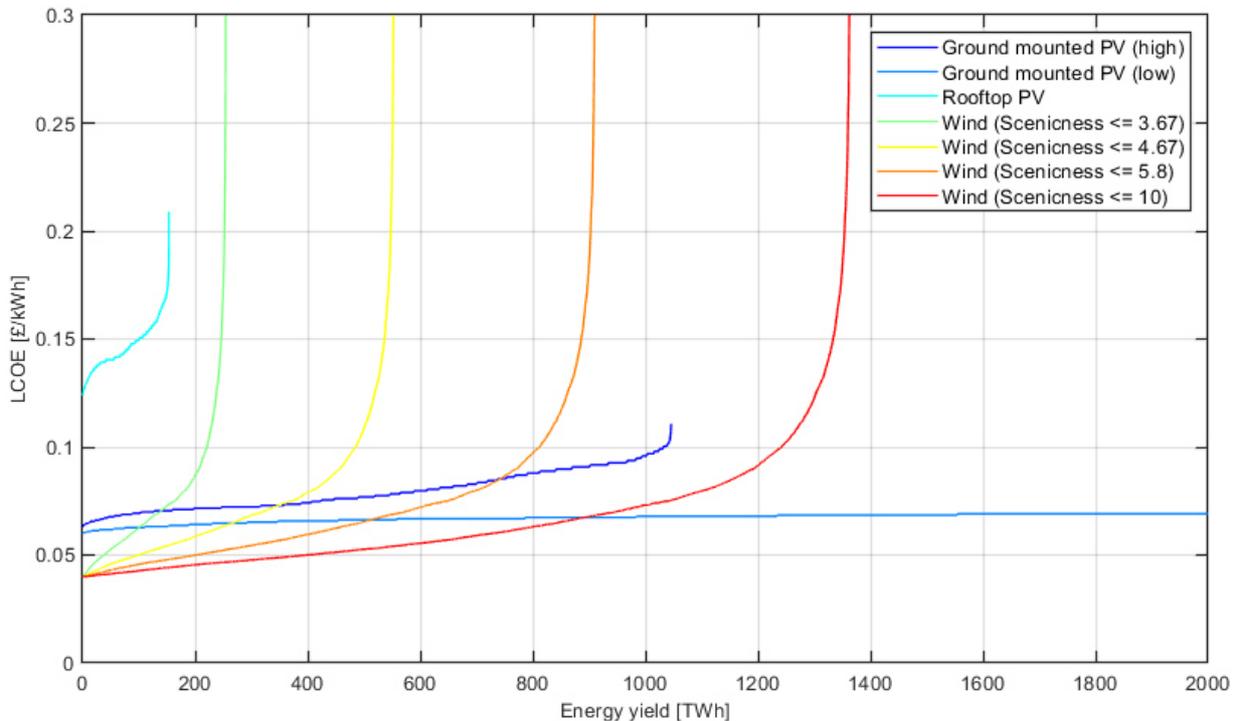

*Figure 2: Cost curves for the three VRE technologies (the curve for ground-mounted PV (high) extends to about 7000 TWh, 0.11 £/kWh). Note the current generation for onshore wind, ground and rooftop PV is about 32, 7 and 5 TWh respectively (BEIS, 2020c, 2014).*

### b. Results at the Local Authority level

This section explores the results at the Local Authority level, based on the classification from 2019 with 382 distinct regions (ONS, 2020). Figure 10 in the Appendix provides an overview of the 382 regions along with their official names, whereas Figure 11 and Figure 12 show their distribution across GB and London respectively. Whilst the area of these regions varies greatly (i.e. from 2 to 26,000 km$^2$), their average size is about 640 km$^2$ – with the exception of seven very large regions, most have areas under 5,000 km$^2$, with 321 under 1,000 km$^2$.

Starting with onshore wind, Figure 3 shows the generation in 2018 (left) alongside the technical potential (middle) and the potential at the 75% scenicness quartile (i.e. <=5.8, right). The data is normalised by the total area of each LA, in GWh/km$^2$, and plotted on two separate logarithmic axes. The left panel of the figure clearly shows the existing distribution of onshore wind generation across GB, with the highest values coinciding with the best wind resource in the south west (e.g. Cornwall and Devon), mid-Wales (e.g. Powys, Ceredigion, Carmarthenshire), northern England (e.g. Northumberland) and most of Scotland, especially the north-west parts. Conversely, the south east and urban areas have relatively low potentials, with the exception of several regions on the east coast and to the north-east of London. The technical potential in the middle panel broadly reflects similar trends, with the contrast between rural and urban locations accentuated.

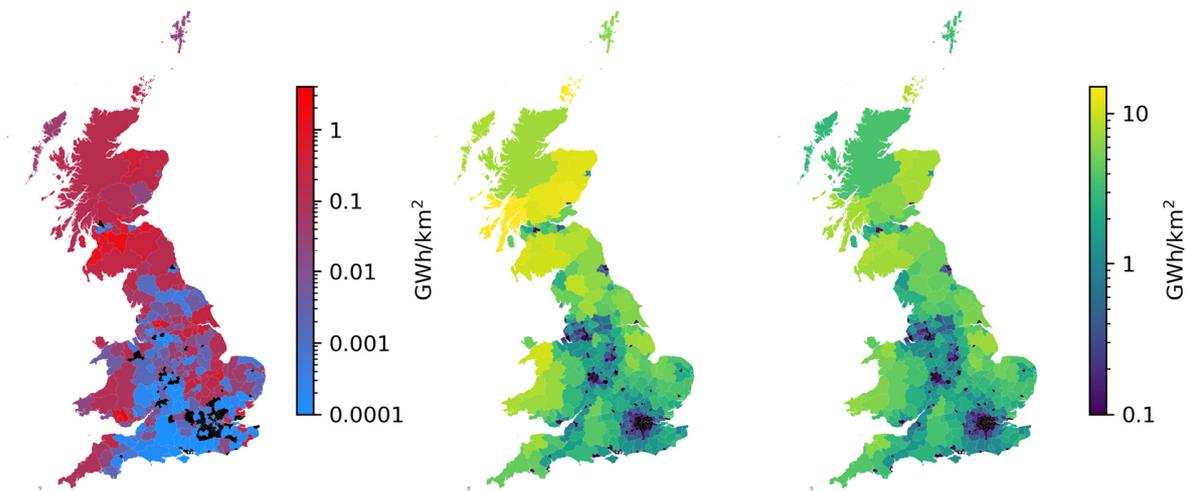

*Figure 3: Onshore wind: generation in 2018 (left, (BEIS, 2020c, 2014)), technical potential (middle) and potential at scenicness <=5.8 (i.e. the 75% quartile, right), in units of GWh/km$^2$. The data is normalised by the total area of each LA and plotted on two separate logarithmic axes.*

Moving to a situation with the 25% most scenic locations excluded, as in the right-hand panel, reveals some interesting differences. In generally, exactly the regions with the largest potentials are those affected by the reduction in available area, confirming that the windiest locations are also the most scenic (McKenna et al., 2021c). The overall potential is reduced and also spread more evenly across the remaining area, with less variation between the regions.

Figure 4 shows the ground-mounted PV generation in 2018 (left), the high restriction scenario (middle) and low restriction scenario (right). Again the data is normalised with the total area of the region and displayed on logarithmic axes. Clear from the left hand panel is the distribution of existing ground-mounted PV systems, which are mainly in lowland areas in England and Wales (black indicates no generation in these figures). Most of Scotland, the mountainous areas of Wales and England, and predominantly urban areas all have little or no generation from this technology. The middle panel (i.e. high restriction scenario) in Figure 4 shows potentials that are still generally higher than the current generation in the left panel, i.e. above 1 GWh/km$^2$. In the right hand panel (low restirction scenario) the technical potential is very high indeed, in many places exceeding 10 GWh/km$^2$. Note that the highest resources for ground-mounted PV partly coincide with those for onshore wind in Figure 3. This will be further explored in section 4.d.

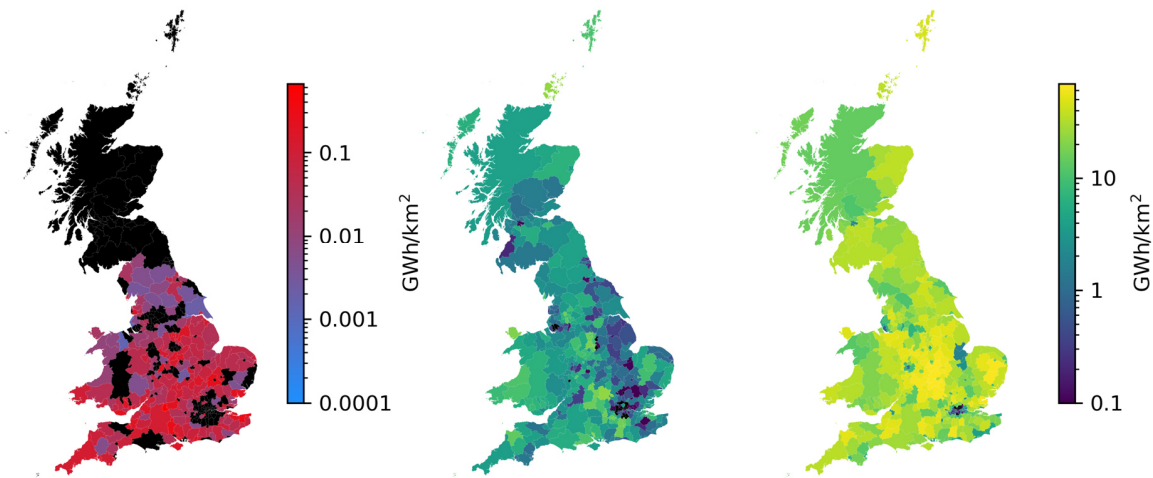

*Figure 4: Ground-mounted solar PV: generation in 2018 (left, (BEIS, 2020c, 2014)), technical potential in high restriction scenario (middle) and low restriction scenario (right) respectively, in units of GWh/km$^2$. The data is normalised by the total area of each LA and plotted on two separate logarithmic axes.*

Figure 5 shows the results for rooftop PV, with the generation in 2018 on the left and the total technical potential on the right. Once more, these results are displayed per unit of the total region's area and scaled on a logarithmic access. The left hand panel illustrates the concentration of existing rooftop PV capacity in primarily urban and/or lowland areas, with only a small number of regions having no installed capacity. The right hand panel demonstrates a substantial remaining potential for this technology, again concentrated in urban areas of GB – clearly visible are London, Birmingham, Manchester, Newcastle and Glasgow.

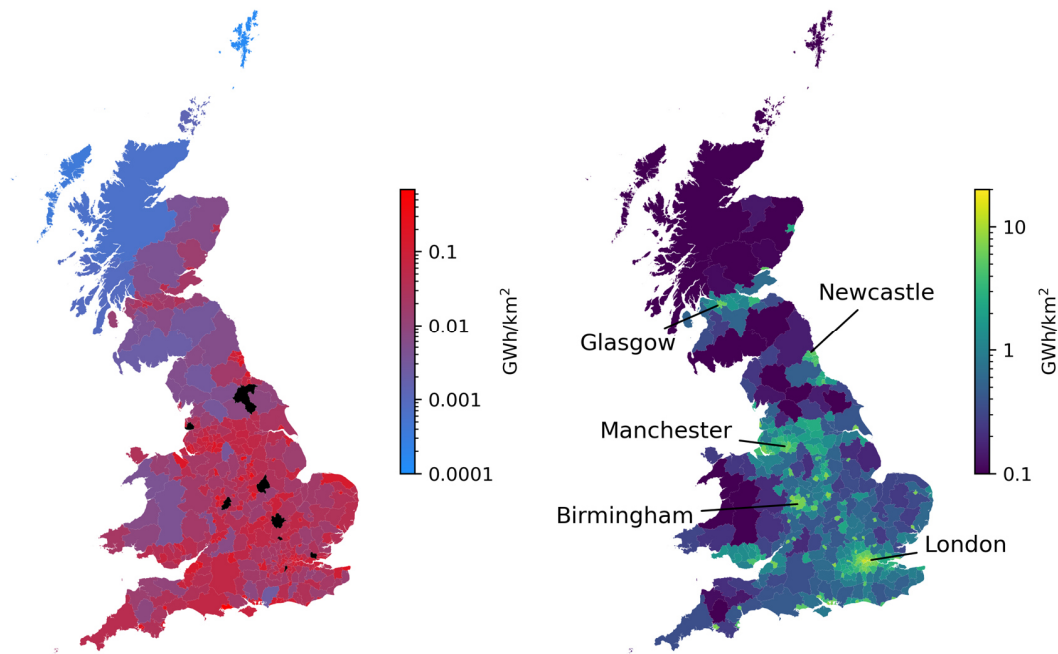

*Figure 5: Rooftop solar PV: generation in 2018 (left, (BEIS, 2020c, 2014)) and total technical potential (right), in units of GWh/km$^2$. The data is normalised by the total area of each LA and plotted on two separate logarithmic axes.*

### c. Scenicness impacts on regional wind potentials and costs

In a parallel study we demonstrated the general link between low-cost wind resources and locations with a high scenicness at the national level (McKenna et al., 2021c). In Figure 6 this relationship is explored for selected Local Authorities, chosen based on the following criteria. Firstly, the regions account for at least 1% each of the total GB onshore wind potential. Secondly, the generation potential in these regions is reduced to at most 98% of this total at scenicness levels up to and including 9. The figure therefore shows one point for each scenicness level from 3 to 10, whereby excluding lower values is due to the very small sample sizes. At each point, the LCOEs should be understood as the mean cumulative ones, i.e. for all scenicness levels up to and including the present one.

It is clear from Figure 6 that there is correlation between the size and quality (as measured by LCOEs) of onshore wind resources in a region and the scenicness values. Removing the locations with the highest scenicness values also removes the lowest cost potential. For some regions, this correlation is stronger than in others – in fact in the Shetland and Orkney Islands the curves are roughly horizontal. This is probably related to an overall very high wind speed/good wind resource throughout these whole regions, and therefore little impact, other than the obvious reduction in potential, when removing the most scenic locations. On the other hand, the locations with the strongest correlation are those such as Eden with the most varied topography and (therefore) wind speeds. In none of these regions does the removal of the most scenic locations reduce the LCOEs.

It seems that this general trend is consistant across all LAs, with regional variations related to land use cover as expected.

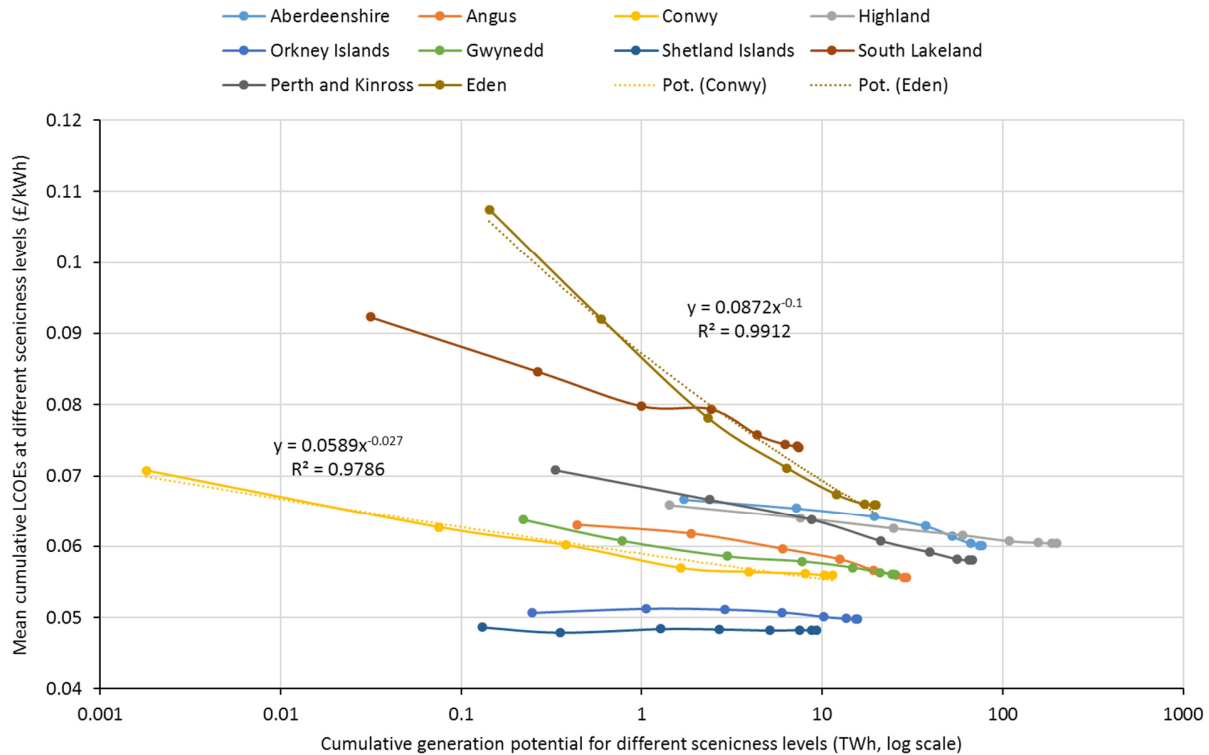

*Figure 6: Mean cumulative LCOEs and cumulative generation potential at discrete scenicness thresholds from 3 to 10 (as shown by points) for selected Local Authority regions in GB (for selection criteria, refer to the text)*

### d. Scenicness and land competition between PV and wind

The combination of PV and wind technologies into hybrid power plants (HPPs) is a well-established concept. WindEurope (2019) identifies diverse motivation for employing these, including optimized network use, high capacity factors, more stable output etc., and review nine examples worldwide. Two types of HPPs are distinguished, namely those where both plants share the same substation and grid connection, and those where the PV panels are integrated with the wind park. The latter are especially relevant in the context of land use competition because they imply a loss in output due to shared land usage. Whilst ground-mounted PV and onshore wind can be closely integrated on the same area of land, their combined capacity density (in MW/km$^2$) is lower than the sum of their individual capacity densities due to required offsets between the technologies. Love (2003) demonstrated this for a highly-renewable US energy system, showing the trade-off between these two technologies when installed to meet a pre-specified demand. In addition, there is a shadowing effect of wind turbines on ground-mounted PV systems, resulting in 1-8% generation losses (Deltenre et al., 2020; Mamia and Appelbaum, 2016).

For these reasons, the overlap between the potentials for onshore wind and ground-mounted PV for selected Local Authorities are shown in Figure 7. These regions are selected based on two criteria, firstly that the overlap at the 75% scenicness threshold (5.8) is at most 80% of the overlap at the 100% threshold, and secondly the overlap in the latter case exceeds 35% of the total region's area. Furthermore, Figure 7 only shows the low restriction PV scenario, as the low PV scenario both exhibits low potentials and generally low overlaps below 20%.

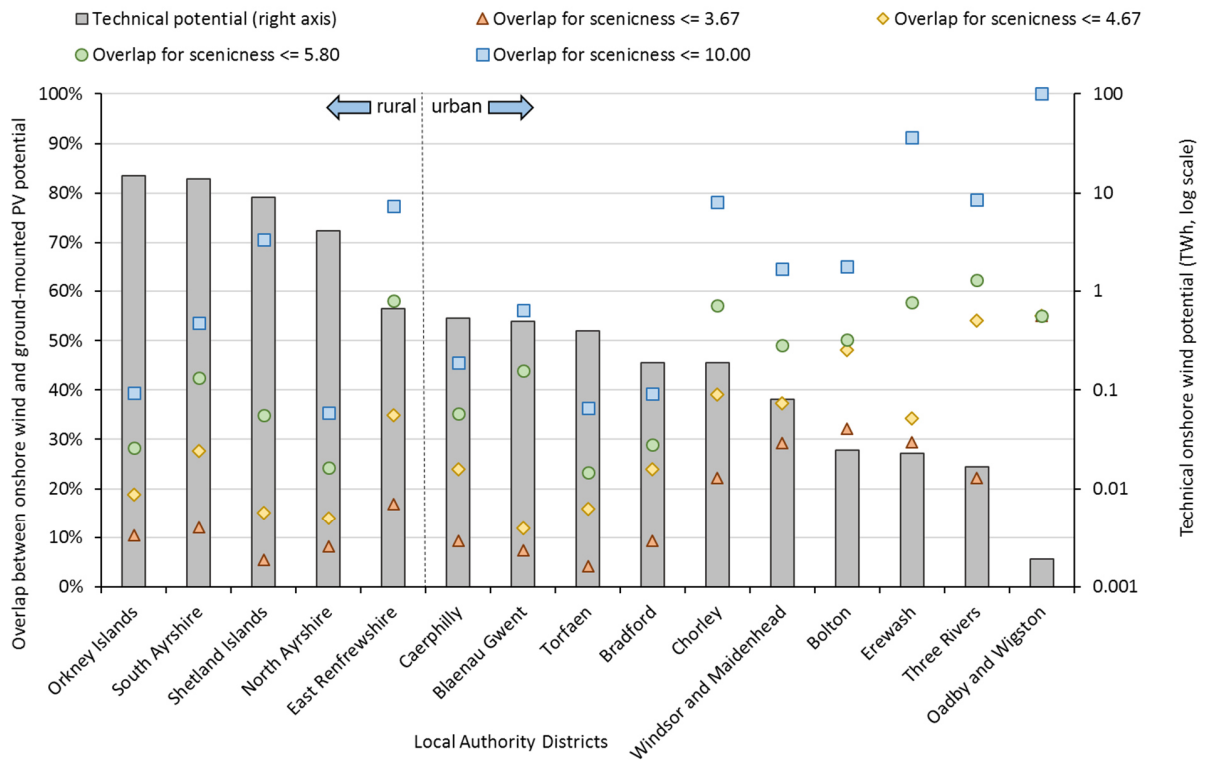

*Figure 7: Overlap between onshore wind and ground-mounted PV at different scenicness values for selected Local Authorities based on criteria detailed in the text (rural and urban denomination is based on predominant land cover category in CORINE (EEA, 2012))*

In the low restriction PV scenario, many mostly urban areas have very high overlaps, as can be seen for the urban areas to the right of Figure 7. The rural regions in this figure are focussed in Scotland (five left-hand bars), where the large area and good wind resource combine to around 45 TWh of generation potential. This is some of the most economic onshore wind potential in GB, located towards the bottom-left of the cost curve in Figure 2. Stepping down the four scenicness thresholds consecutively but differently reduces the overall generation potential and overlap in these regions. In rural regions the overlap varies from about 5-80%, whereas the urban regions show a range of about 10-100%. Overall, the areas with the best onshore wind resource have some of the largest overlaps, meaning inevitable trade-offs between technologies and criteria in the context of constrained land availability.

## 5. Critical discussion and validation of methodology

One of the main novelties in the presented method is the integration of bottom-up with top-down approaches to rooftop PV estimations. Whilst the BU method in Mainzer et al. (2017) has already been validated, the extension developed here has not been. This section is devoted to the validation of the new hybrid top-down/bottom-up method for rooftop PV potential estimation, beginning with a case study for the city of Leeds and followed by a comparison of the results for the whole Yorkshire and Humber region, including 15 LA districts.

In order to give a deeper insight into the results and to validate these with existing plants and studies, we focus on Leeds as one of the cities employed to link bottom-up and top-down approaches. Figure 7 shows the results of the bottom-up method in four panels, whereby a detailed map of city with wards can be found in the Appendix. Panel a) shows the locations of partial rooftop areas and existing PV plants (from Stowell et al. (2020)), both displayed as point clouds, with a cutout of several collocated points in east Leeds. Panel b) shows the CLC land use categories for the same area, whereas panels c) and d) plot the technical potential (in kWh/m$^2$) and LCOEs (£/kWh) respectively.

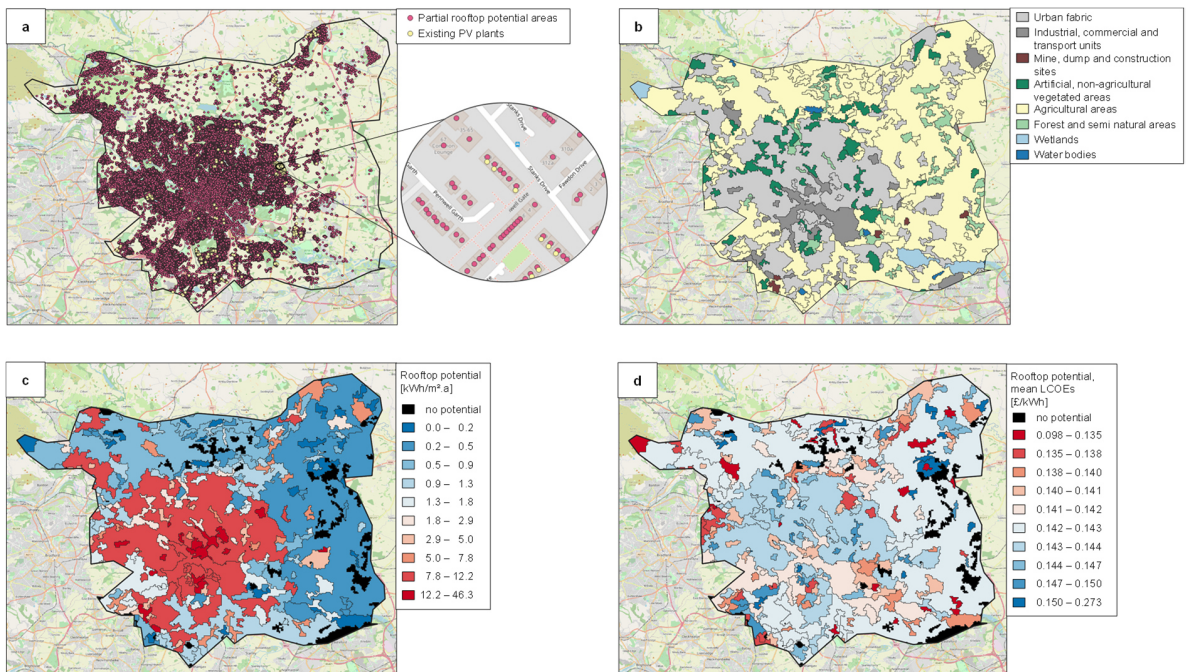

*Figure 8: Results for PV rooftop and existing plants in Leeds: a) locations of partial rooftop areas and existing PV plants (from Stowell et al. (2020)), b) CLC land use categories c) technical potential (in kWh/m2) and d) LCOEs (£/kWh)*

Despite several studies having assessed the national potential for onshore wind and rooftop PV (Table 7), relatively few peer-reviewed studies analyze individual cities. Many studies are undertaken on a consultancy basis and are published as technical reports. In addition to those studies cited in Table 7**Fehler! Verweisquelle konnte nicht gefunden werden.**, for example, the Mayor Of London has published a Solar Action Plan which targets 2 GW of rooftop PV by 2050, but the Plan does not include any detailed resource assessment (Greater London Authority, 2018). In addition, Calderdale Metropolitan Borough Council (2018) published a Renewable and Low Carbon Energy Plan including details of renewable resources. One limitation of these one-off studies for individual cities is that the methodology often differs, which makes comparisons challenging. Against this background, an extensive study by AECOM (2011) covering large parts of Yorkshire and the Humber is particularly relevant. This report details resource assessments for the whole portfolio of renewable energy (heat and power) technologies across multiple Local Authorities in the region. For this reason of broad coverage, this source is employed here for the purpose of validation.

      The AECOM (2011) study assumes for domestic buildings that 25% of the existing stock and 50% of new build developments represent technically accessible resources. Due to the low rate of new build (typically 1-2%) the average here is nearer to 25% than 50% (giving a factor of 2-4 compared to the technical potential). Further, commercial and industrial buildings in the existing stock are assumed to be 40% and 80% usable respectively, with more modest assumptions of 5-30% for new builds. In a second step, module sizes are predefined for these three building types of 2 kW, 5 kW and 10 kW on domestic, commercial and industrial buildings respectively (this results in a further factor of 2-3 compared to the technical potential). In a third step, the economically viable resource is assessed based on assumed proportions of the building stock: 5-40% in 2010 and 18-45% in 2016 (this step introduces a factor of 2-10 relative to the technical potential). In order to make the results comparable to this paper, the results from AECOM (2011) are therefore multiplied by a factor of eight (i.e. the product of the previously-mentioned factors).

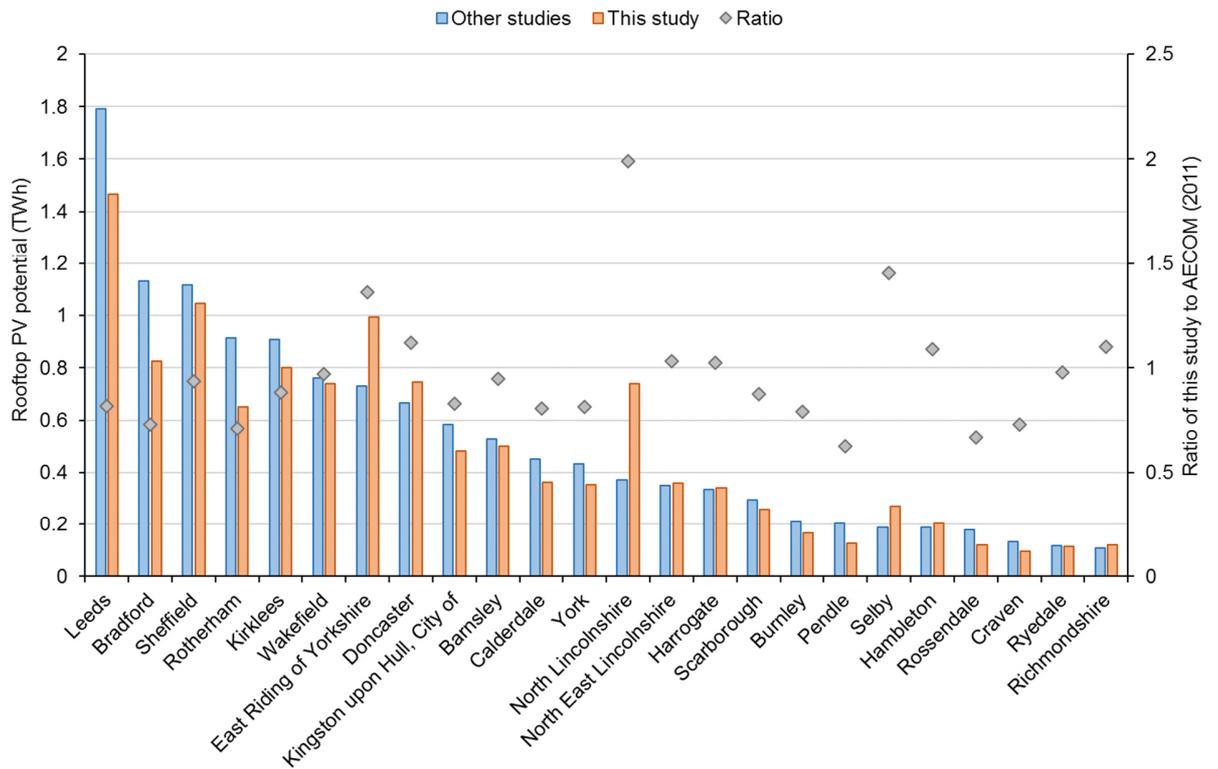

Figure 9 shows the results of the present paper for rooftop PV in comparison to those from AECOM (2011), whereby the left axis displays the generation potential and the right axes specifies the ratio between this study and the other study. Whilst there is clearly a wide variation in the rooftop-PV potentials in different cities/regions, the agreement between the two sources is reasonably good. In other words, the ratio of the two is close to unity in most cases, with a mean of 0.97 and a standard deviation 0.30).

Despite this overall generally good agreement between the two studies, there are some large deviations in individual cases. In order to explore this phenomenon, we analysed the land use distributions in each of the 24 cities shown in

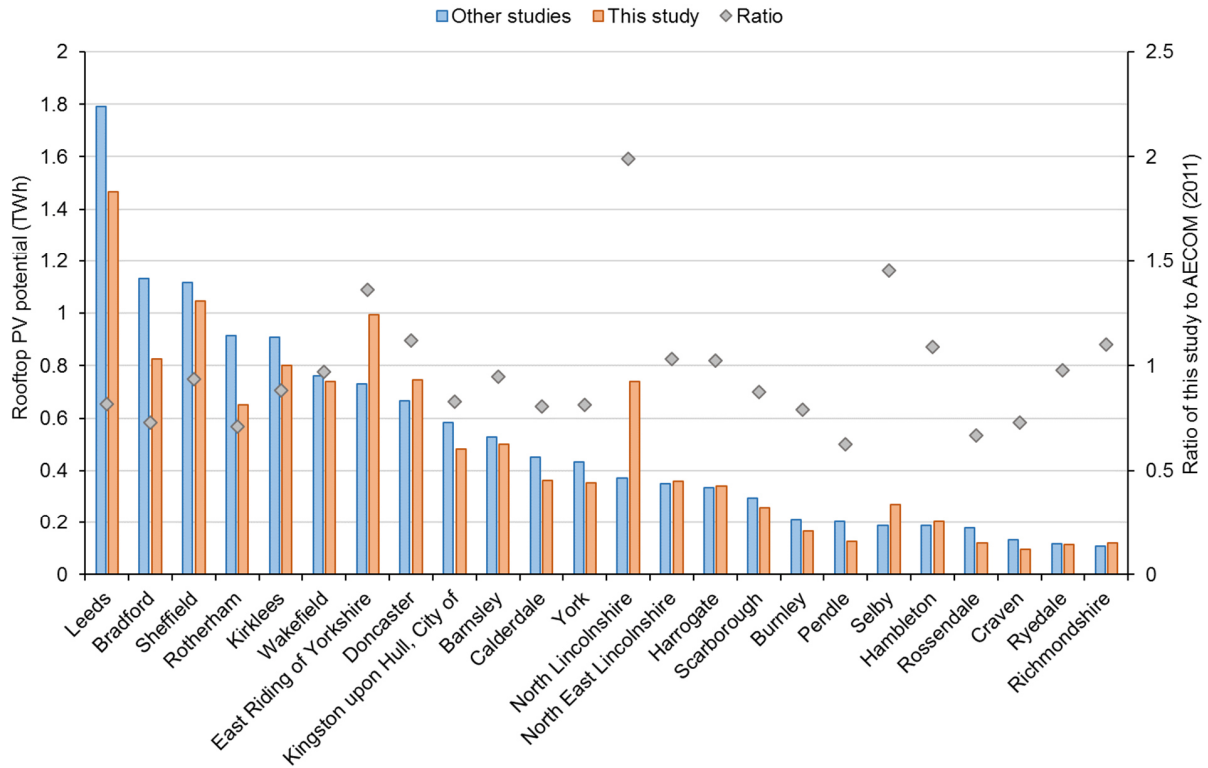

Figure 9. Due to having 13 land use categories (MHCLG, 2020) and only 24 observations (cities), we aggregated land use categories in order to reduce the number of model coefficients, as shown in Equation 7 below. Table 5 shows the descriptive statistics for this dataset.

*Vacant land = Undeveloped land + Vacant*
*Commercial land use = CommunityService + IndustryandCommerce + Defense Buildings*
*Other = Unknown developed use + Minerals and landfill + Transport and utilities + Outdoor recreation* (7)
*Agricultural land and forest = Agriculture + Forest open land and water + ResidentialGardens*

Table 5: Descriptive statistics for land use dataset and 24 cities

|  | Mean | Std. Dev. | Min. | Max. |
|---|---|---|---|---|
| Deviation | 0.97 | 0.30 | 0.62 | 1.99 |
| Residential land (%) | 1.85 | 1.79 | 0.15 | 8.43 |
| Commercial land use (%) | 1.99 | 2.06 | 0.12 | 10.01 |
| Vacant land (%) | 1.73 | 1.67 | 0.08 | 7.82 |
| Agriculture land and forests (%) | 83.26 | 13.21 | 34.99 | 97.16 |
| Other land use | 11.17 | 7.76 | 2.41 | 38.74 |

Notes: Deviation is defined as the ratio between the results of the present paper for rooftop PV in comparison to those from AECOM (2011); Commercial land use includes community service buildings, industry and commerce buildings and defense buildings; vacant land includes underdeveloped land and vacant land; other land use includes outdoor recreation facilities, transport infrastructure, landfills and unknown use. Number of observations is 24.

It should be noted here that, by definition, the sum of land use shares for each city is 100. This means that coefficients for all land use categories (N) cannot be estimated, but only for N-1. Hence we employ the land use category "vacant" as the basis and present the regression results in Table 6.

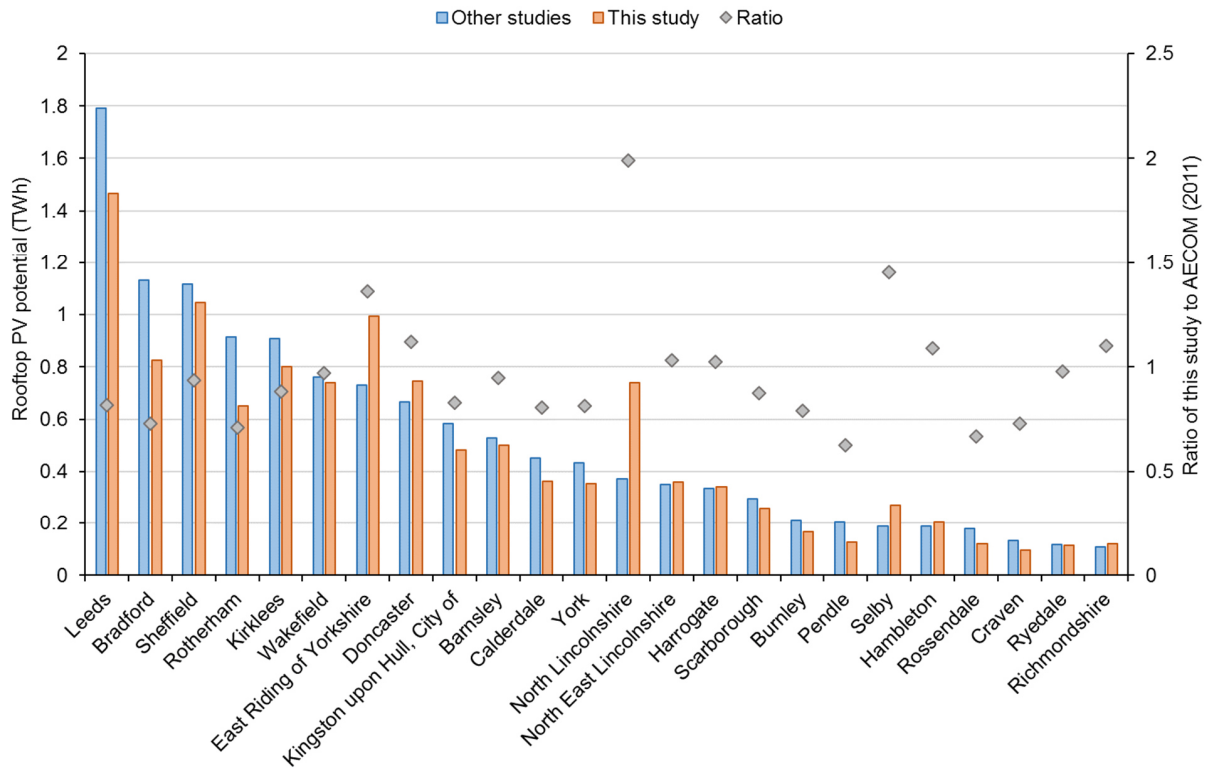

*Figure 9: Comparison of results of this study for rooftop PV with those of AECOM (2011) (left hand axis) and the ratio of the two (right hand axis)*

The interpretation of these results is straightforward. Firstly, it should be noted that only the coefficient associated with residential land use is significant. The coefficient associated with commercial land use is only borderline significant. The other coefficients are not significantly different from zero. The estimation results therefore suggest that a higher share of the residential land use is associated with a lower deviation (ratio) between the study results. In other words, the proportion of residential land use is a strongly influencing factor for the correspondence (or otherwise) between the two methods.

Here we also briefly put the results of this paper into the context of previous studies with a similar focus. The lack of research into the potential for ground-mounted PV in a national context means that we focus on onshore wind and rooftop PV. Table 7 below shows these potentials from eleven other sources, whereby not all cover both technologies. It should also be noted that the scope differs slightly between the UK, GB and the British Isles, as does the potential assessed, between technical and feasible.

The range of results for onshore wind in terms of technical potential are 1274-4700 TWh for the UK, and the range for rooftop PV is 44-540 TWh. In both cases, the present study, with an estimated 1324 TWh and 153 TWh for wind and rooftop PV respectively, lies well within and towards the lower end of this range. Certainly for the rooftop PV, the results it closer to the studies for feasible potential[1], especially the recent one from Vivid economics and ICL (2019).

Table 6: Regression results for the deviation variable defined in the text

|  | (1) Deviation |
|---|---|
| Residential land (%) | -0.878*** |
|  | (0.262) |
| Commercial land use (%) | 0.637** |
|  | (0.258) |
| Agriculture land and forests (%) | 0.017 |
|  | (0.160) |
| Other land use (%) | 0.053 |
|  | (0.178) |
| Constant | -0.681 |
|  | (16.07) |
| Number of observations | 24 |
| $R^2$ | 0.595 |

Notes: Deviation is defined as the ratio between the results of the present paper for rooftop PV in comparison to those from AECOM (2011); commercial land use includes community service buildings, industry and commerce buildings and defense buildings; vacant land includes underdeveloped land and vacant land; other land use includes outdoor recreation facilities, transport infrastructure, landfills and unknown use. Vacant land use is the basis land use category. Standard errors are in parentheses. ** $p<0.05$, *** $p<0.01$.

Hence the results can be interpreted as well within the range of existing studies and rather towards the conservative end for both technologies, notwithstanding minor differences in geographical coverage. Further discussion of underlying reasons for differences in results can be found in McKenna et al. (2020).

Table 7: Potentials for onshore wind and rooftop PV from selected studies

| Study | Onshore wind potential (TWh) | Rooftop PV (TWh) | Geographical scope | Potential definition |
|---|---|---|---|---|
| Bódis et al. (2019) |  | 44 | UK | Technical |
| UK PMA (2009) |  | 460 | UK | Technical |
| Defaix et al. (2012) |  | 80 | UK | Technical |
| ETSU (1999) | 318 | 266 (BIPV[#]) | UK | Total accessible |
| Vivid economics and ICL (2019) | 215-479[$] | 35[$] | GB | Feasible |
| MacKay (2008) | 4700 | 115* | UK | Technical |
| Dalla Longa et al. (2018) | 1391 |  | UK | Technical |

| Enevoldsen et al. (2019) | 2302 | | British Isles | 'Socio-technical' |
|---|---|---|---|---|
| Ryberg et al. (2019) | 2262 | | UK | Technical |
| EEA (2009) | 3961-4409 | | UK | Technical |
| (McKenna et al., 2015) | 1274 | | UK | Technical |
| This study | 1324 | 153 | GB | Technical |

\* South facing roofs only; # Building-integrated PV; $ Based on 37 GW with 950 h FLH and 96-214 GW with 2240 h FLH for rooftop PV and wind respectively (BEIS, 2020b).

At this point it should also be pointed out that the developed method is based on some simplifying assumptions and is therefore not intended as a replacement for detailed local resource assessments with better quality data. Indeed, assumptions such as that all modules are facing south and relying on 83% of the population living in cities mean that the method has weaknesses in specific (e.g. rural) locations. But the key advantage in this method is in the broad coverage, as it relies solely on open data that is widely available for other countries. In principle, the method is highly transferable to any location in the world where both OSM and Bing maps have reasonable coverage, and a land use database such as CORINE or equivalent is available.

## 6. Conclusions and policy implications

In this concluding section, we reflect on some of the current issues with RE planning across GB and highlight some of the emergent tensions between energy and planning policy, before proposing five recommendations for realigning the two in the context of net zero.

British planning policy and the implementation has a significant influence on individual energy projects and thus energy pathways more broadly. RE projects are subject to decision making across multiple scales and arenas of governance. At the national level, divergence of planning responsibilities has resulted in a patchwork of approaches across GB's devolved administrations. Scotland has remained supportive of onshore wind with the 2017 Onshore Wind Policy Statement (Scottish Government, 2017) and net zero commitments are embedded in the 2020 4th National Planning Framework (Scottish Government, 2020). A spatial framework serves to highlight those areas most (and least) likely to gain approval based on National Park or National Scenic Area designations. Also broadly supportive of onshore wind, Wales has since 2005 considered wind proposals in the context of seven 'Strategic Search Areas', which are most appropriate for onshore wind (Welsh Government, 2005). The 2021 National Development Framework (Welsh Government, 2021) supersedes this, replacing area-based targets with a national target. It includes a presumption in favour of large-scale wind energy developments, subject to some constraints. Onshore wind proposals in England, however, have since 2015 been heavily restricted by the National Planning Policy Framework, which requires projects to be aligned with wind provision set out in Local or Neighbourhood Plans, as well as demonstrate additional community backing at the point of application (Smith, 2016).

At the individual project level, planning decisions for VREs are shaped by local contexts and politics. The existence of multiple (often conflicting) stakeholder interests means that energy projects and pathways are shaped by conflict and negotiation (Bridge et al., 2013; Roelich and Giesekam, 2019). While this may be the case across the whole of GB, contentious projects are more likely to find approval when decisions can be defaulted to an overarching national policy, as is the case for onshore wind in Scotland. Elsewhere, there is a need for pragmatism to reconcile the interests and values of actors (Bhardwaj et al., 2019).

As a relatively incoherent patchwork of policy statements, spatial approaches and governance arrangements, the GB energy planning landscape has arguably failed to evolve in step with political and cultural attitudes towards RE technologies. While public support for onshore wind at a national level has increased significantly over the last decade, onshore wind has only really found support in Scotland. However, new net zero commitments at the UK level will require planning regimes that are much more coherent with energy policy across all jurisdictions in order to provide an enabling environment for local energy developments (CCC, 2020b; HoC, 2019).

The apparent trade-offs between good locations for VRE technologies, scenic landscapes and other land uses discussed in this paper suggest the need for realignment between planning policy and energy policy across local and national scales. Five key issues/recommendations for research and policy can be highlighted in this regard.

First, while some trade-offs are inevitable, having an overarching national strategic vision for land use across GB embedded within planning regimes can provide clarity for developers and decision-makers alike. Planning has evolved from being underpinned by the notion of the 'public interest', and more recently around the similarly ambiguous objective of 'sustainable development' (Maidment, 2016). Given the specific trade-offs highlighted here for scenic landscapes, and interdependencies between land use and energy pathways (e.g. bioenergy cropping, forestation), there is a strong argument now for planning to be strategically aligned explicitly with net zero objectives (CCC, 2020a). This would not eradicate trade-offs of course, but could lend coherence in favouring decisions that provide efficient and just GHG mitigation impacts.

Second, meeting the net zero challenge requires a significant increase in VRE penetration and it is generally agreed that a diverse portfolio of technologies will be needed to maximise overall RE deployment and reduce the need for additional flexibility (CCC, 2019; PÖYRY, 2011). In this context, robust and transparent appraisal of the synergies and trade-offs between development options – alongside other land uses – will be needed to legitimize support for specific technologies as well as the decision-making processes adopted (Rohe and Chlebna, 2021; Smith, 2007).

A third consideration stems from the increased emphasis among academics, policymakers, and innovation agencies on the importance of local contexts as a key part of a whole system approach to decarbonisation. The Energy Systems Catapult (2018) for example suggests that spatial planning of low carbon developments should be considered (alongside energy network planning and demand-side regulations) within an integrated Local Area Energy Planning framework. Understanding the potential for trade-offs across different localities will be an important consideration in such frameworks.

Fourth, it is apparent that decarbonising energy is increasingly a challenge of technological integration, rather than *only* deployment of VREs. As such, decision-making around proposed RE projects needs to account for any impacts projects may have on the electricity system, e.g. the costs of balancing supply and demand, or the need to constrain or store excess VRE generation. The decision-making around trade-offs and synergies (through co-location of different technologies, for example) therefore need to take place in the context of such whole-system cost assessments (BEIS, 2020a). Future spatial modelling work in this space should also seek to move away from levelised cost of technologies as a basis for understanding trade-offs. Some inroads in this direction have been made in some parallel related work to this article (McKenna et al., 2021c; Price et al., 2020).

Finally, the quality of decision making at any level of governance will be determined by the degree to which relevant interests can be taken into account. Such interests include the value placed on scenic landscapes – as discussed here – although other factors are also likely to play significant roles. In order for local RE development to respond to the climate change mitigation imperative, more meaningful engagement with the public is needed, particularly in those areas where these potential trade-offs are strongest. This could most readily take the form of encouraging best practice (and clarifying the meaning thereof) around community engagement as a necessary component of RE project proposals (regen and Electricity Storage Network, 2020). Examples of this best practice here include promotion of shared ownership, inclusion of community-led organisations and wider communities throughout all project stages (rather than just the planning stage), and maximisation of local employment opportunities.

More generally, however, the development of national and local climate assemblies in the UK offer replicable frameworks for public deliberation around climate change responses (Climate Assembly UK, 2021; Mellier-Wilson and Toy, 2020). Such fora provide valuable mechanisms for opening up discussions about GHG mitigation options, as well as the trade-offs these options might have with environmental and social outcomes.

## 7. Acknowledgements

The authors gratefully acknowledge the contributions of David Schlund, who carried out some of the wind analysis whilst a Student Assistant at KIT, as well as Camille Moutard, upon whose Master Thesis at DTU this article builds (*Assessing the 'acceptable' onshore wind potential in the UK*, 2019, https://findit.dtu.dk/en/catalog/2451029061). The usual disclaimer applies.

# 8. Appendices

A list and maps of the Local Authority regions employed in this analysis is shown in Figures 10, 11 and 12 below.

| # | Name | # | Name | # | Name | # | Name | # | Name | # | Name |
|---|---|---|---|---|---|---|---|---|---|---|---|
| 1 | Hartlepool | 65 | South Cambridgeshire | 129 | Canterbury | 193 | Rushcliffe | 257 | Trafford | 320 | Belfast |
| 2 | Middlesbrough | 66 | Allerdale | 130 | Dartford | 194 | Cherwell | 258 | Wigan | 321 | Causeway Coast and Glens |
| 3 | Redcar and Cleveland | 67 | Barrow-in-Furness | 131 | Dover | 195 | Oxford | 259 | Knowsley | 322 | Derry City and Strabane |
| 4 | Stockton-on-Tees | 68 | Carlisle | 132 | Gravesham | 196 | South Oxfordshire | 260 | Liverpool | 323 | Fermanagh and Omagh |
| 5 | Darlington | 69 | Copeland | 133 | Maidstone | 197 | Vale of White Horse | 261 | St. Helens | 324 | Lisburn and Castlereagh |
| 6 | Halton | 70 | Eden | 134 | Sevenoaks | 198 | West Oxfordshire | 262 | Sefton | 325 | Mid and East Antrim |
| 7 | Warrington | 71 | South Lakeland | 135 | Folkestone and Hythe | 199 | Mendip | 263 | Wirral | 326 | Mid Ulster |
| 8 | Blackburn with Darwen | 72 | Amber Valley | 136 | Swale | 200 | Sedgemoor | 264 | Barnsley | 327 | Newry, Mourne and Down |
| 9 | Blackpool | 73 | Bolsover | 137 | Thanet | 201 | South Somerset | 265 | Doncaster | 328 | Ards and North Down |
| 10 | Kingston upon Hull, City of | 74 | Chesterfield | 138 | Tonbridge and Malling | 202 | Cannock Chase | 266 | Rotherham | 329 | Clackmannanshire |
| 11 | East Riding of Yorkshire | 75 | Derbyshire Dales | 139 | Tunbridge Wells | 203 | East Staffordshire | 267 | Sheffield | 330 | Dumfries and Galloway |
| 12 | North East Lincolnshire | 76 | Erewash | 140 | Burnley | 204 | Lichfield | 268 | Newcastle upon Tyne | 331 | East Ayrshire |
| 13 | North Lincolnshire | 77 | High Peak | 141 | Chorley | 205 | Newcastle-under-Lyme | 269 | North Tyneside | 332 | East Lothian |
| 14 | York | 78 | North East Derbyshire | 142 | Fylde | 206 | South Staffordshire | 270 | South Tyneside | 333 | East Renfrewshire |
| 15 | Derby | 79 | South Derbyshire | 143 | Hyndburn | 207 | Stafford | 271 | Sunderland | 334 | Na h-Eileanan Siar |
| 16 | Leicester | 80 | East Devon | 144 | Lancaster | 208 | Staffordshire Moorlands | 272 | Birmingham | 335 | Falkirk |
| 17 | Rutland | 81 | Exeter | 145 | Pendle | 209 | Tamworth | 273 | Coventry | 336 | Highland |
| 18 | Nottingham | 82 | Mid Devon | 146 | Preston | 210 | Babergh | 274 | Dudley | 337 | Inverclyde |
| 19 | Herefordshire, County of | 83 | North Devon | 147 | Ribble Valley | 211 | Ipswich | 275 | Sandwell | 338 | Midlothian |
| 20 | Telford and Wrekin | 84 | South Hams | 148 | Rossendale | 212 | Mid Suffolk | 276 | Solihull | 339 | Moray |
| 21 | Stoke-on-Trent | 85 | Teignbridge | 149 | South Ribble | 213 | Elmbridge | 277 | Walsall | 340 | North Ayrshire |
| 22 | Bath and North East Somerset | 86 | Torridge | 150 | West Lancashire | 214 | Epsom and Ewell | 278 | Wolverhampton | 341 | Orkney Islands |
| 23 | Bristol, City of | 87 | West Devon | 151 | Wyre | 215 | Guildford | 279 | Bradford | 342 | Scottish Borders |
| 24 | North Somerset | 88 | Eastbourne | 152 | Blaby | 216 | Mole Valley | 280 | Calderdale | 343 | Shetland Islands |
| 25 | South Gloucestershire | 89 | Hastings | 153 | Charnwood | 217 | Reigate and Banstead | 281 | Kirklees | 344 | South Ayrshire |
| 26 | Plymouth | 90 | Lewes | 154 | Harborough | 218 | Runnymede | 282 | Leeds | 345 | South Lanarkshire |
| 27 | Torbay | 91 | Rother | 155 | Hinckley and Bosworth | 219 | Spelthorne | 283 | Wakefield | 346 | Stirling |
| 28 | Swindon | 92 | Wealden | 156 | Melton | 220 | Surrey Heath | 284 | Gateshead | 347 | Aberdeen City |
| 29 | Peterborough | 93 | Basildon | 157 | North West Leicestershire | 221 | Tandridge | 285 | City of London | 348 | Aberdeenshire |
| 30 | Luton | 94 | Braintree | 158 | Oadby and Wigston | 222 | Waverley | 286 | Barking and Dagenham | 349 | Argyll and Bute |
| 31 | Southend-on-Sea | 95 | Brentwood | 159 | Boston | 223 | Woking | 287 | Barnet | 350 | City of Edinburgh |
| 32 | Thurrock | 96 | Castle Point | 160 | East Lindsey | 224 | North Warwickshire | 288 | Bexley | 351 | Renfrewshire |
| 33 | Medway | 97 | Chelmsford | 161 | Lincoln | 225 | Nuneaton and Bedworth | 289 | Brent | 352 | West Dunbartonshire |
| 34 | Bracknell Forest | 98 | Colchester | 162 | North Kesteven | 226 | Rugby | 290 | Bromley | 353 | West Lothian |
| 35 | West Berkshire | 99 | Epping Forest | 163 | South Holland | 227 | Stratford-on-Avon | 291 | Camden | 354 | Angus |
| 36 | Reading | 100 | Harlow | 164 | South Kesteven | 228 | Warwick | 292 | Croydon | 355 | Dundee City |
| 37 | Slough | 101 | Maldon | 165 | West Lindsey | 229 | Adur | 293 | Ealing | 356 | East Dunbartonshire |
| 38 | Windsor and Maidenhead | 102 | Rochford | 166 | Breckland | 230 | Arun | 294 | Enfield | 357 | Fife |
| 39 | Wokingham | 103 | Tendring | 167 | Broadland | 231 | Chichester | 295 | Greenwich | 358 | Perth and Kinross |
| 40 | Milton Keynes | 104 | Uttlesford | 168 | Great Yarmouth | 232 | Crawley | 296 | Hackney | 359 | Glasgow City |
| 41 | Brighton and Hove | 105 | Cheltenham | 169 | King's Lynn and West Norfolk | 233 | Horsham | 297 | Hammersmith and Fulham | 360 | North Lanarkshire |
| 42 | Portsmouth | 106 | Cotswold | 170 | North Norfolk | 234 | Mid Sussex | 298 | Haringey | 361 | Isle of Anglesey |
| 43 | Southampton | 107 | Forest of Dean | 171 | Norwich | 235 | Worthing | 299 | Harrow | 362 | Gwynedd |
| 44 | Isle of Wight | 108 | Gloucester | 172 | South Norfolk | 236 | Bromsgrove | 300 | Havering | 363 | Conwy |
| 45 | County Durham | 109 | Stroud | 173 | Corby | 237 | Malvern Hills | 301 | Hillingdon | 364 | Denbighshire |
| 46 | Cheshire East | 110 | Tewkesbury | 174 | Daventry | 238 | Redditch | 302 | Hounslow | 365 | Flintshire |
| 47 | Cheshire West and Chester | 111 | Basingstoke and Deane | 175 | East Northamptonshire | 239 | Worcester | 303 | Islington | 366 | Wrexham |
| 48 | Shropshire | 112 | East Hampshire | 176 | Kettering | 240 | Wychavon | 304 | Kensington and Chelsea | 367 | Ceredigion |
| 49 | Cornwall | 113 | Eastleigh | 177 | Northampton | 241 | Wyre Forest | 305 | Kingston upon Thames | 368 | Pembrokeshire |
| 50 | Isles of Scilly | 114 | Fareham | 178 | South Northamptonshire | 242 | St Albans | 306 | Lambeth | 369 | Carmarthenshire |
| 51 | Wiltshire | 115 | Gosport | 179 | Wellingborough | 243 | Welwyn Hatfield | 307 | Lewisham | 370 | Swansea |
| 52 | Bedford | 116 | Hart | 180 | Craven | 244 | East Hertfordshire | 308 | Merton | 371 | Neath Port Talbot |
| 53 | Central Bedfordshire | 117 | Havant | 181 | Hambleton | 245 | Stevenage | 309 | Newham | 372 | Bridgend |
| 54 | Northumberland | 118 | New Forest | 182 | Harrogate | 246 | East Suffolk | 310 | Redbridge | 373 | Vale of Glamorgan |
| 55 | Bournemouth, Christchurch and Poole | 119 | Rushmoor | 183 | Richmondshire | 247 | West Suffolk | 311 | Richmond upon Thames | 374 | Cardiff |
| 56 | Dorset | 120 | Test Valley | 184 | Ryedale | 248 | Somerset West and Taunton | 312 | Southwark | 375 | Rhondda Cynon Taf |
| 57 | Aylesbury Vale | 121 | Winchester | 185 | Scarborough | 249 | Bolton | 313 | Sutton | 376 | Caerphilly |
| 58 | Chiltern | 122 | Broxbourne | 186 | Selby | 250 | Bury | 314 | Tower Hamlets | 377 | Blaenau Gwent |
| 59 | South Bucks | 123 | Dacorum | 187 | Ashfield | 251 | Manchester | 315 | Waltham Forest | 378 | Torfaen |
| 60 | Wycombe | 124 | Hertsmere | 188 | Bassetlaw | 252 | Oldham | 316 | Wandsworth | 379 | Monmouthshire |
| 61 | Cambridge | 125 | North Hertfordshire | 189 | Broxtowe | 253 | Rochdale | 317 | Westminster | 380 | Newport |
| 62 | East Cambridgeshire | 126 | Three Rivers | 190 | Gedling | 254 | Salford | 318 | Antrim and Newtownabbey | 381 | Powys |
| 63 | Fenland | 127 | Watford | 191 | Mansfield | 255 | Stockport | 319 | Armagh City, Banbridge and Craigavon | 382 | Merthyr Tydfil |
| 64 | Huntingdonshire | 128 | Ashford | 192 | Newark and Sherwood | 256 | Tameside | | | | |

*Figure 10: List of LA regions and codes employed in Figure 11 (ONS, 2020)*

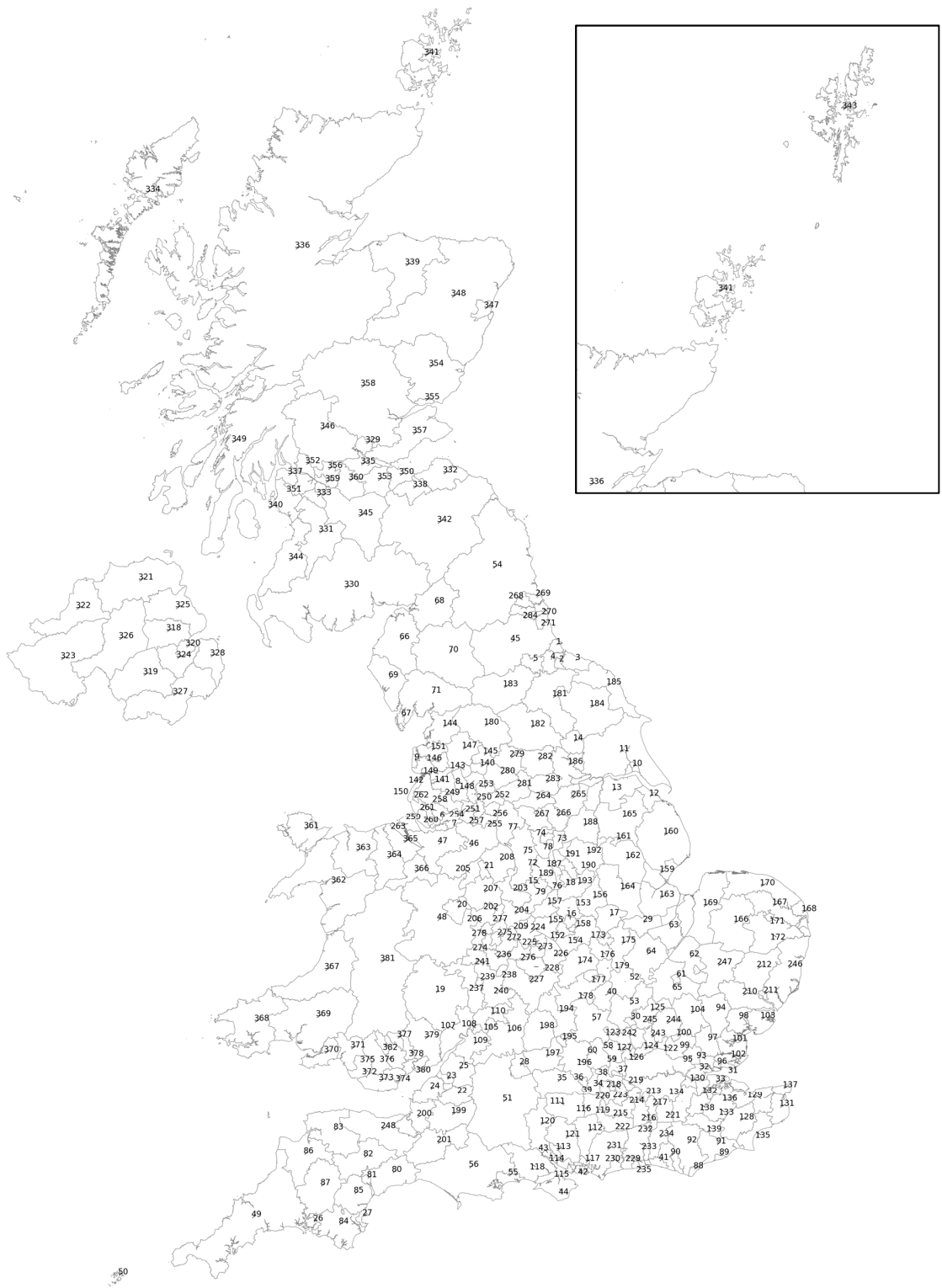

Figure 11: Map of UK showing numbered LA regions corresponding to the list in Figure 10 (the Shetland Islands are cut off in the north-east corner of the map due to space constraints) (ONS, 2020)

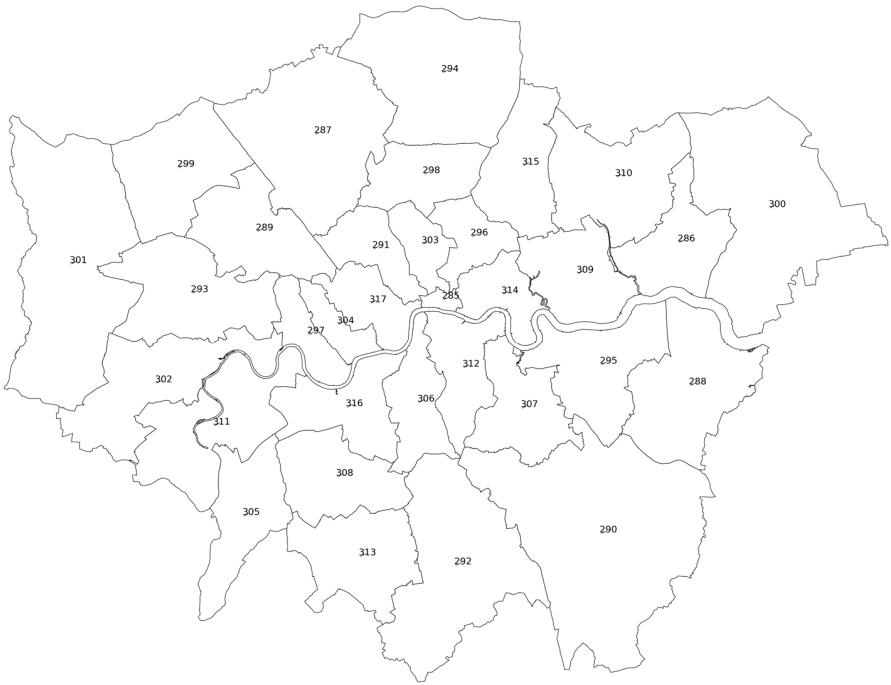

*Figure 12: Detail of LA regions in Greater London, corresponding to those in the list in Figure 10 (ONS, 2020)*

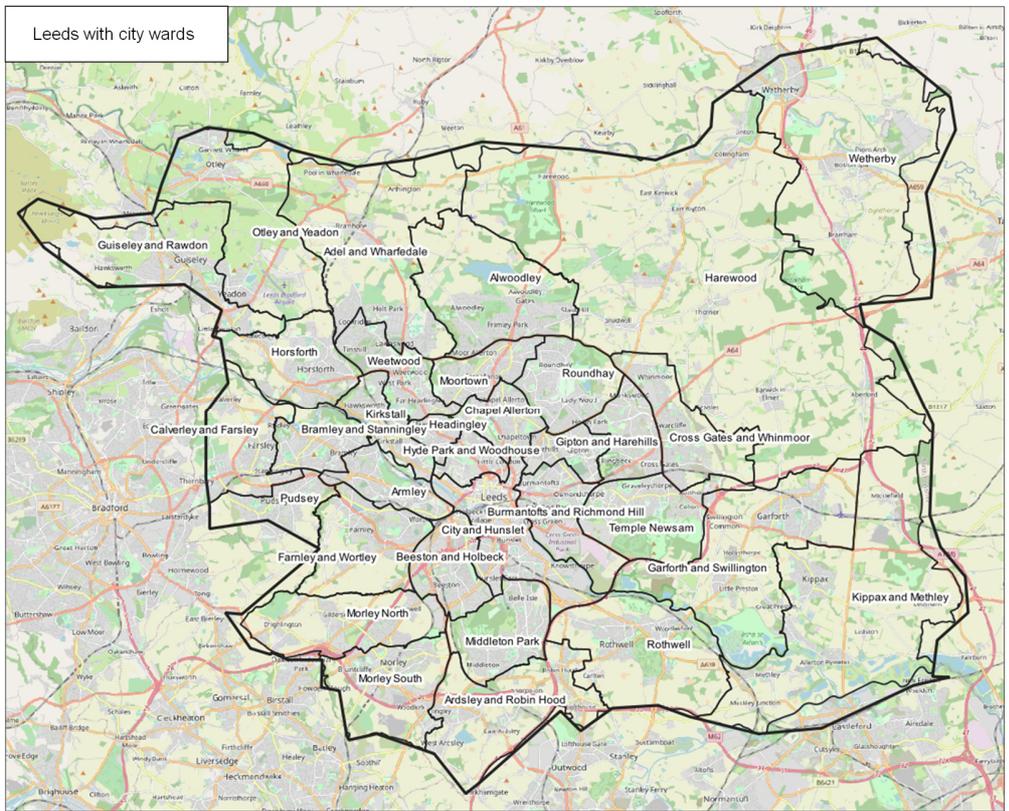

*Figure 13: Map of Leeds city showing city wards for comparison with Figure 7 in the main text*

## 9. References

AECOM, 2011. Low carbon and renewable energy capacity in Yorkshire and Humber. https://www.kirklees.gov.uk/beta/planning-policy/pdf/supportingDocuments/climateChange/Low-Carbon-Renewable-Energy-Capacity-Yorkshire-Humber.pdf (accessed 17 March 2021).

Barclay, C., 2012. Wind Farms - Distance from housing. https://unidoc.wiltshire.gov.uk/UniDoc/Document/File/MTQvMDg3NzgvRlVMLDU4MDk5Nw== (accessed 22 March 2021).

BEIS, 2014. Renewable Energy Planning Database quarterly extract. Renewable Energy Planning Database quarterly extract (accessed 1 March 2021).

BEIS, 2018. Energy and Climate Change Public Attitude Tracker: WAVE 25. https://www.gov.uk/government/statistics/energy-and-climate-change-public-attitudes-tracker-wave-25 (accessed 1 March 2021).

BEIS, 2020a. BEIS Electricity Generation Costs (2020). https://www.gov.uk/government/publications/beis-electricity-generation-costs-2020 (accessed 7 March 2021).

BEIS, 2020b. Digest of UK Energy Statistics (DUKES): renewable sources of energy. https://www.gov.uk/government/statistics/renewable-sources-of-energy-chapter-6-digest-of-united-kingdom-energy-statistics-dukes (accessed 7 March 2021).

BEIS, 2020c. Renewable electricity by local authority. https://assets.publishing.service.gov.uk/government/uploads/system/uploads/attachment_data/file/920656/Renewable_electricity_by_local_authority_2014_to_2019.xlsx (accessed 1 March 2021).

BEIS, 2020d. Solar photovoltaic (PV) cost data. https://www.gov.uk/government/statistics/solar-pv-cost-data (accessed 17 March 2021).

BEIS, 2020e. Total final energy consumption at regional and local authority level: 2005 to 2018. https://www.gov.uk/government/statistics/total-final-energy-consumption-at-regional-and-local-authority-level-2005-to-2018 (accessed 1 March 2021).

Betakova, V., Vojar, J., Sklenicka, P., 2015. Wind turbines location: How many and how far? Applied Energy 151, 23–31. https://doi.org/10.1016/j.apenergy.2015.04.060.

Bhardwaj, A., Joshi, M., Khosla, R., Dubash, N.K., 2019. More priorities, more problems? Decision-making with multiple energy, development and climate objectives. Energy Research & Social Science 49, 143–157. https://doi.org/10.1016/j.erss.2018.11.003.

Bódis, K., Kougias, I., Jäger-Waldau, A., Taylor, N., Szabó, S., 2019. A high-resolution geospatial assessment of the rooftop solar photovoltaic potential in the European Union. Renewable and Sustainable Energy Reviews 114, 109309. https://doi.org/10.1016/j.rser.2019.109309.

Borgogno Mondino, E., Fabrizio, E., Chiabrando, R., 2015. Site Selection of Large Ground-Mounted Photovoltaic Plants: A GIS Decision Support System and an Application to Italy. International Journal of Green Energy 12, 515–525. https://doi.org/10.1080/15435075.2013.858047.

Boudet, H.S., 2019. Public perceptions of and responses to new energy technologies. Nat Energy 4, 446–455. https://doi.org/10.1038/s41560-019-0399-x.

Bridge, G., Bouzarovski, S., Bradshaw, M., Eyre, N., 2013. Geographies of energy transition: Space, place and the low-carbon economy. Energy Policy 53, 331–340. https://doi.org/10.1016/j.enpol.2012.10.066.

Calderdale Metropolitan Borough Council, 2018. Calderdale local plan: Renewable and low carbon energy. https://www.calderdale.gov.uk/v2/sites/default/files/Local-plan-eb-renewable-low-carbon-tp_2018.pdf (accessed 17 March 2021).

Carrión, J.A., Espín Estrella, A., Aznar Dols, F., Ridao, A.R., 2008. The electricity production capacity of photovoltaic power plants and the selection of solar energy sites in Andalusia (Spain). Renewable Energy 33, 545–552. https://doi.org/10.1016/j.renene.2007.05.041.

CCC, 2019. Net Zero - Technical Annex: Integrating variable renewables into the UK electricity system. https://www.theccc.org.uk/wp-content/uploads/2019/05/Net-Zero-Technical-Annex-Integrating-variable-renewables.pdf (accessed 13 May 2021).

CCC, 2020a. Land use: Policies for Net Zero UK. https://www.theccc.org.uk/publication/land-use-policies-for-a-net-zero-uk/ (accessed 13 May 2021).

CCC, 2020b. The Sixth Carbon Budget: The UK`s path to Net Zero. https://www.theccc.org.uk/wp-content/uploads/2020/12/The-Sixth-Carbon-Budget-The-UKs-path-to-Net-Zero.pdf (accessed 7 March 2021).

Climate Assembly UK, 2021. The path to net zero. https://www.climateassembly.uk/report/read/index.html (accessed 13 May 2021).

Dalla Longa, F., Kober, T., Badger, J., Volker, P., Hoyer-Klick, C., Hidalgo Gonzalez, I., Medarac, H., Nijs, W., Politis, S., Tardvydas, D., Zucker, A., 2018. Wind potentials for EU and neighbouring countries: Input datasets for the JRC-EU-TIMES Model. https://ec.europa.eu/jrc/en/publication/wind-potentials-eu-and-neighbouring-countries-input-datasets-jrc-eu-times-model (accessed 7 March 2021).

DCLG, 2017. Final Decision on Scout Moor Wind Farm Application. https://assets.publishing.service.gov.uk/government/uploads/system/uploads/attachment_data/file/625856/17-07-06_FINAL_DL_Scout_Moor_Wind_Farm.pdf (accessed 1 March 2021).

Defaix, P.R., van Sark, W.G.J.H.M., Worrell, E., Visser, E. de, 2012. Technical potential for photovoltaics on buildings in the EU-27. Solar Energy 86, 2644–2653. https://doi.org/10.1016/j.solener.2012.06.007.

DEFRA, 2019. Rural population 2014/15. https://www.gov.uk/government/statistics/rural-population-and-migration/rural-population-201415 (accessed 1 March 2021).

Deltenre, Q., Troyer, T. de, Runacres, M.C., 2020. Performance assessment of hybrid PV-wind systems on high-rise rooftops in the Brussels-Capital Region. Energy and Buildings 224, 110137. https://doi.org/10.1016/j.enbuild.2020.110137.

EEA, 2009. Europe's onshore and offshore wind energy potential. https://www.eea.europa.eu/publications/europes-onshore-and-offshore-wind-energy-potential (accessed 7 March 2021).

EEA, 2012. CORINE Land Cover 2012. https://www.eea.europa.eu/data-and-maps/data/external/corine-land-cover-2012 (accessed 1 March 2021).

Energy Systems Catapult, 2018. Local Area Energy Planning: Guidance for local authorities and energy providers. https://es.catapult.org.uk/brochures/local-area-energy-planning-guidance-for-local-authorities-and-energy-providers/ (accessed 13 May 2021).

Enevoldsen, P., Permien, F.-H., Bakhtaoui, I., Krauland, A.-K. von, Jacobson, M.Z., Xydis, G., Sovacool, B.K., Valentine, S.V., Luecht, D., Oxley, G., 2019. How much wind power potential does europe have? Examining european wind power potential with an enhanced socio-technical atlas. Energy Policy 132, 1092–1100. https://doi.org/10.1016/j.enpol.2019.06.064.

ETSU, 1999. New and Renewable Energy: Prospects in the UK for the 21st Century: Supporting Analysis. https://books.google.de/books/about/New_and_Renewable_Energy.html?id=Kz9qjwEACAAJ&redir_esc=y (accessed 7 March 2021).

European Commission, 2018. SARAH Solar Radiation Data. https://ec.europa.eu/jrc/en/PVGIS/downloads/SARAH (accessed 1 March 2021).

European Commission, 2019. Photovoltaic Geographical Information System. https://re.jrc.ec.europa.eu/pvg_tools/en/ (accessed 17 March 2021).

Fast, S., Mabee, W., Baxter, J., Christidis, T., Driver, L., Hill, S., McMurtry, J.J., Tomkow, M., 2016. Lessons learned from Ontario wind energy disputes. Nat Energy 1. https://doi.org/10.1038/nenergy.2015.28.

Gangopadhyay, U., Jana, S., Das, S., 2013. State of Art of Solar Photovoltaic Technology. Conference Papers in Energy 2013, 1–9. https://doi.org/10.1155/2013/764132.

Gassar, A.A.A., Cha, S.H., 2021. Review of geographic information systems-based rooftop solar photovoltaic potential estimation approaches at urban scales. Applied Energy 291, 116817. https://doi.org/10.1016/j.apenergy.2021.116817.

Greater London Authority, 2018. Solar Action Plan for London. https://www.london.gov.uk/WHAT-WE-DO/environment/environment-publications/solar-action-plan (accessed 17 March 2021).

Harper, M., Anderson, B., James, P., Bahaj, A., 2019. Assessing socially acceptable locations for onshore wind energy using a GIS-MCDA approach. International Journal of Low-Carbon Technologies 14, 160–169. https://doi.org/10.1093/ijlct/ctz006.

HoC, 2019. Clean Growth: Technologies for meeting the UK's emissions reduction targets. https://publications.parliament.uk/pa/cm201719/cmselect/cmsctech/1454/1454.pdf (accessed 13 May 2021).

Höltinger, S., Salak, B., Schauppenlehner, T., Scherhaufer, P., Schmidt, J., 2016. Austria's wind energy potential – A participatory modeling approach to assess socio-political and market acceptance. Energy Policy 98, 49–61. https://doi.org/10.1016/j.enpol.2016.08.010.

Hoogwijk, M., 2004. On the Global and Regional Potential of Renewable Energy. https://dspace.library.uu.nl/bitstream/handle/1874/782/full.pdf (accessed 1 March 2021).

Jäger, T., McKenna, R., Fichtner, W., 2016. The feasible onshore wind energy potential in Baden-Württemberg: A bottom-up methodology considering socio-economic constraints. Renewable Energy 96, 662–675. https://doi.org/10.1016/j.renene.2016.05.013.

JNCC, 2016. UK Protected Area Datasets for Download. https://jncc.gov.uk/our-work/uk-protected-area-datasets-for-download/ (accessed 7 March 2021).

Konadu, D.D., Mourão, Z.S., Allwood, J.M., Richards, K.S., Kopec, G., McMahon, R., Fenner, R., 2015. Land use implications of future energy system trajectories—The case of the UK 2050 Carbon Plan. Energy Policy 86, 328–337. https://doi.org/10.1016/j.enpol.2015.07.008.

Liddiard, R., 2021. Personal correspondance.

Love, M., 2003. Land Area and Storage Requirements for Wind and Solar Generation to Meet the US Hourly Electrical Demand. Dissertation.

MacKay, D.J.C., 2008. Sustainable Energy - without the hot air. https://www.withouthotair.com/ (accessed 7 March 2021).

Maidment, C., 2016. In the public interest? Planning in the Peak District National Park. Planning Theory 15, 366–385. https://doi.org/10.1177/1473095216662093.

Mainzer, K., Fath, K., McKenna, R., Stengel, J., Fichtner, W., Schultmann, F., 2014. A high-resolution determination of the technical potential for residential-roof-mounted photovoltaic systems in Germany. Solar Energy 105, 715–731. https://doi.org/10.1016/j.solener.2014.04.015.

Mainzer, K., Killinger, S., McKenna, R., Fichtner, W., 2017. Assessment of rooftop photovoltaic potentials at the urban level using publicly available geodata and image recognition techniques. Solar Energy 155, 561–573. https://doi.org/10.1016/j.solener.2017.06.065.

Mamia, I., Appelbaum, J., 2016. Shadow analysis of wind turbines for dual use of land for combined wind and solar photovoltaic power generation. Renewable and Sustainable Energy Reviews 55, 713–718. https://doi.org/10.1016/j.rser.2015.11.009.

McFadden, 1974. Conditional Logit Analysis of Qualitative Choice Behavior, in: Zarembka, P. (Ed.), Frontiers in econometrics. Academic Press, New York, pp. 105–142.

McKenna, R., Hollnaicher, S., Fichtner, W., 2014. Cost-potential curves for onshore wind energy: A high-resolution analysis for Germany. Applied Energy 115, 103–115. https://doi.org/10.1016/j.apenergy.2013.10.030.

McKenna, R., Hollnaicher, S., v. d. Ostman Leye, P., Fichtner, W., 2015. Cost-potentials for large onshore wind turbines in Europe. Energy 83, 217–229. https://doi.org/10.1016/j.energy.2015.02.016.

McKenna, R., Petrovic, S., Weinand, J., Mulalic, I., Mainzer, K., Price, J., Soutar, I., 2021a. Exploring trade-offs between landscape impact, land use and resource quality for onshore variable renewable energy: an application to Great Britain. figshare. Dataset. https://figshare.com/articles/dataset/Exploring_trade-offs_between_landscape_impact_land_use_and_resource_quality_for_onshore_variable_renewable_energy_an_application_to_Great_Britain/14595864 (accessed 14 May 2021).

McKenna, R., Pfenninger, S., Heinrichs, H., Schmidt, J., Staffell, I., Gruber, K., Hahmann, A.N., Jansen, M., Klingler, M., Landwehr, N., Larsén, X.G., Lilliestam, J., Pickering, B., Robinius,

M., Tröndle, T., Turkovska, O., Wehrle, S., Weinand, J.M., Wohland, J., 2021b. Reviewing methods and assumptions for high-resolution large-scale onshore wind energy potential assessments. https://arxiv.org/pdf/2103.09781.

McKenna, R., Ryberg, D.S., Staffell, I., Hahmann, A.N., Schmidt, J., Heinrichs, H., Höltinger, S., Lilliestam, J., Pfenninger, S., Robinius, M., Stolten, D., Tröndle, T., Wehrle, S., Weinand, J.M., 2020. On the socio-technical potential for onshore wind in Europe: A response to Enevoldsen et al. (2019), Energy Policy, 132, 1092-1100. Energy Policy 145, 111693. https://doi.org/10.1016/j.enpol.2020.111693.

McKenna, R., Weinand, J.M., Mulalic, I., Petrovic, S., Mainzer, K., Preis, T., Moat, H.S., 2021c. Scenicness assessment of onshore wind sites with geotagged photographs and impacts on approval and cost-efficiency. Nat Energy In press.

Mellier-Wilson, C., Toy, S., 2020. UK Climate Change Citizens' Assemblies & Citizens' Juries. https://www.involve.org.uk/resources/case-studies/uk-climate-change-citizens-assemblies-citizens-juries (accessed 13 May 2021).

MHCLG, 2020. Live tables on land use. https://www.gov.uk/government/statistical-data-sets/live-tables-on-land-use (accessed 17 March 2021).

Ministry of Agriculture, Fisheries and Food, 1988. Agricultural Land Classification detailed Post 1988 ALC survey, Cranfield, Home Farm (Mid Beds LP Site 7) (ALCC01595B). http://publications.naturalengland.org.uk/publication/6487172215996416 (accessed 1 March 2021).

Molnarova, K., Sklenicka, P., Stiborek, J., Svobodova, K., Salek, M., Brabec, E., 2012. Visual preferences for wind turbines: Location, numbers and respondent characteristics. Applied Energy 92, 269–278. https://doi.org/10.1016/j.apenergy.2011.11.001.

Office for National Statistics, 2020. National Parks (August 2016) Boundaries GB BFE. https://data.gov.uk/dataset/cd98fbc9-cbde-46ef-bf2a-7c21852be01f/national-parks-august-2016-boundaries-gb-bfe (accessed 1 March 2021).

ofgem, 2021. Bills, prices and profits. https://www.ofgem.gov.uk/publications-and-updates/infographic-bills-prices-and-profits (accessed 17 March 2021).

Ong, S., Campbell, C., Denholm, P., Margolis, R., Heath, G., 2013. Land-Use Requirements for Solar Power Plants in the United States. https://www.nrel.gov/docs/fy13osti/56290.pdf (accessed 1 March 2021).

ONS, 2020. Local Authority Districts (December 2019) Boundaries UK BFC. https://geoportal.statistics.gov.uk/datasets/1d78d47c87df4212b79fe2323aae8e08_0 (accessed 7 March 2021).

Open Energy Information, 2017. Transparent Cost Database. https://openei.org/apps/TCDB/transparent_cost_database (accessed 7 March 2021).

OpenStreetMap Contributors, 2020. OpenStreetMap. https://www.openstreetmap.org/#map=7/56.188/11.617 (accessed 1 March 2021).

Ordnance Survey, 2018. OS Terrain 50. https://osdatahub.os.uk/downloads/open/Terrain50 (accessed 1 March 2021).

Ordnance Survey, 2021. Batch coordinate transformation tool. https://www.ordnancesurvey.co.uk/gps/transformation/batch (accessed 1 March 2021).

Perpiña Castillo, C., Batista e Silva, F., Lavalle, C., 2016. An assessment of the regional potential for solar power generation in EU-28. Energy Policy 88, 86–99. https://doi.org/10.1016/j.enpol.2015.10.004.

Petrova, M.A., 2016. From NIMBY to acceptance: Toward a novel framework — VESPA — For organizing and interpreting community concerns. Renewable Energy 86, 1280–1294. https://doi.org/10.1016/j.renene.2015.09.047.

Philipps, S., Warmuth, W., 2019. Photovoltaics Report. https://www.ise.fraunhofer.de/de/veroeffentlichungen/studien/photovoltaics-report.html (accessed 7 March 2021).

PÖYRY, 2011. Analysing technical constraints on renewable generation to 2050. https://afry.com/sites/default/files/2020-11/analysing-the-technical-constraints-on-renewable-generation.pdf (accessed 13 May 2021).

Price, J., Mainzer, K., Petrovic, S., Zeyringer, M., McKenna, R., 2020. The implications of landscape visual impact on future highly renewable power systems: a case study for Great Britain. IEEE Trans. Power Syst., 1. https://doi.org/10.1109/TPWRS.2020.2992061.

Price, J., Zeyringer, M., Konadu, D., Sobral Mourão, Z., Moore, A., Sharp, E., 2018. Low carbon electricity systems for Great Britain in 2050: An energy-land-water perspective. Applied Energy 228, 928–941. https://doi.org/10.1016/j.apenergy.2018.06.127.

regen, Electricity Storage Network, 2020. Contracts for Difference (CFD): Proposed Amendments to the Scheme 2020. https://www.regen.co.uk/wp-content/uploads/Regen-response-to-CFD-consultation-May-2020-v1.0.pdf (accessed 13 May 2021).

Roddis, P., Carver, S., Dallimer, M., Norman, P., Ziv, G., 2018. The role of community acceptance in planning outcomes for onshore wind and solar farms: An energy justice analysis. Applied Energy 226, 353–364. https://doi.org/10.1016/j.apenergy.2018.05.087.

Roelich, K., Giesekam, J., 2019. Decision making under uncertainty in climate change mitigation: introducing multiple actor motivations, agency and influence. Climate Policy 19, 175–188. https://doi.org/10.1080/14693062.2018.1479238.

Rohe, S., Chlebna, C., 2021. A spatial perspective on the legitimacy of a technological innovation system: Regional differences in onshore wind energy. Energy Policy 151, 112193. https://doi.org/10.1016/j.enpol.2021.112193.

Ryberg, D.S., Caglayan, D.G., Schmitt, S., Linßen, J., Stolten, D., Robinius, M., 2019. The future of European onshore wind energy potential: Detailed distribution and simulation of advanced turbine designs. Energy 182, 1222–1238. https://doi.org/10.1016/j.energy.2019.06.052.

Schumacher, K., Krones, F., McKenna, R., Schultmann, F., 2019. Public acceptance of renewable energies and energy autonomy: A comparative study in the French, German and Swiss Upper Rhine region. Energy Policy 126, 315–332. https://doi.org/10.1016/j.enpol.2018.11.032.

Scotland's soil, 1981. National scale land capability for agriculture at a scale of 1:250000. https://soils.environment.gov.scot/maps/capability-maps/national-scale-land-capability-for-agriculture/ (accessed 1 March 2021).

Scottish Government, 2017. Onshore Wind Policy Statement. https://www.gov.scot/binaries/content/documents/govscot/publications/speech-statement/2017/12/onshore-wind-policy-statement-9781788515283/documents/00529536-pdf/00529536-pdf/govscot%3Adocument/00529536.pdf (accessed 13 May 2021).

Scottish Government, 2020. Scotland's Fourth National Planning Framework Position Statement. https://www.gov.scot/binaries/content/documents/govscot/publications/progress-report/2020/11/scotlands-fourth-national-planning-framework-position-statement/documents/scotlands-fourth-national-planning-framework-position-statement/scotlands-fourth-national-planning-framework-position-statement/govscot%3Adocument/scotlands-fourth-national-planning-framework-position-statement.pdf?forceDownload=true (accessed 13 May 2021).

Silva, J., Riberiro, C., Guedes, R., 2007. Roughness Length classification of Corine Land Cover Classes. https://citeseerx.ist.psu.edu/viewdoc/download?doi=10.1.1.608.2707&rep=rep1&type=pdf (accessed 22 March 2021).

Smith, A., 2007. Emerging in between: The multi-level governance of renewable energy in the English regions. Energy Policy 35, 6266–6280. https://doi.org/10.1016/j.enpol.2007.07.023.

Smith, L., 2016. Planning for onshore wind. https://researchbriefings.files.parliament.uk/documents/SN04370/SN04370.pdf (accessed 13 May 2021).

Sonnberger, M., Ruddat, M., 2017. Local and socio-political acceptance of wind farms in Germany. Technology in Society 51, 56–65. https://doi.org/10.1016/j.techsoc.2017.07.005.

Stowell, D., Kelly, J., Tanner, D., Taylor, J., Jones, E., Geddes, J., Chalstrey, E., 2020. A harmonised, high-coverage, open dataset of solar photovoltaic installations in the UK. Scientific data 7, 394. https://doi.org/10.1038/s41597-020-00739-0.

UK Government, 2008. Climate Change Act 2008. https://www.legislation.gov.uk/ukpga/2008/27/pdfs/ukpga_20080027_en.pdf (accessed 7 March 2021).

UK Met Office, 2018. Weather and Climate data. https://www.metoffice.gov.uk/services/data (accessed 22 March 2021).

UK PMA, 2009. A vision for UK PV. Cited in Rudd, H. (2015): A literature survey of UK renewables potential. https://www.howardrudd.net/files/RenewablesPotential.pdf (accessed 7 March 2021).

van der Horst, D., 2007. NIMBY or not? Exploring the relevance of location and the politics of voiced opinions in renewable energy siting controversies. Energy Policy 35, 2705–2714. https://doi.org/10.1016/j.enpol.2006.12.012.

Vartiainen, E., Masson, G., Breyer, C., Moser, D., Román Medina, E., 2020. Impact of weighted average cost of capital, capital expenditure, and other parameters on future utility-scale PV levelised cost of electricity. Prog Photovolt Res Appl 28, 439–453. https://doi.org/10.1002/pip.3189.

Vivid economics, ICL, 2019. Accelerated Electrification and the GB Electricity System. https://www.theccc.org.uk/publication/accelerated-electrification-and-the-gb-electricity-system/ (accessed 7 March 2021).

Warren, C.R., Lumsden, C., O'Dowd, S., Birnie, R.V., 2005. 'Green On Green': Public perceptions of wind power in Scotland and Ireland. Journal of Environmental Planning and Management 48, 853–875. https://doi.org/10.1080/09640560500294376.

Welsh Government, 2005. Planning Policy Wales: Technical Advice Notice: 8: Planning for Renewable Energy. https://gov.wales/sites/default/files/publications/2018-09/tan8-renewable-energy_0.pdf (accessed 13 May 2021).

Welsh Government, 2017. Predictive Agricultural Land Classification (ALC) Map. https://lle.gov.wales/catalogue/item/PredictiveAgriculturalLandClassificationALCMap/?lang=en (accessed 1 March 2021).

Welsh Government, 2021. Future Wales: The National Plan 2040. https://gov.wales/sites/default/files/publications/2021-02/future-wales-the-national-plan-2040.pdf (accessed 13 May 2021).

WindEurope, 2019. Renewable Hybrid Power Plants: Exploring the benefits and market opportunities.

Wolsink, M., 2018. Co-production in distributed generation: renewable energy and creating space for fitting infrastructure within landscapes. Landscape Research 43, 542–561. https://doi.org/10.1080/01426397.2017.1358360.

YouGov, 2018. Renewable UK Survey. http://d25d2506sfb94s.cloudfront.net/cumulus_uploads/document/3hx70b1nzc/RenewableUK_June18_GB_w.pdf (accessed 1 March 2021).